\documentclass[a4paper,12pt]{article}
\pdfoutput=1
\usepackage{jheppub}
\usepackage{amsfonts}
\usepackage{slashed}
\usepackage{subfigure}
\usepackage{array}
\usepackage{verbatim}
\usepackage{mathrsfs}
\usepackage{graphicx}
\usepackage{dcolumn}
\usepackage{bm}
\usepackage{amsmath}
\usepackage{amssymb}
\usepackage{multirow}
\usepackage{epsfig,epstopdf}
\usepackage{cancel}

\def\lsim{\mathrel{\raise.3ex\hbox{$<$\kern-.75em\lower1ex\hbox{$\sim$}}}}
\def\gsim{\mathrel{\raise.3ex\hbox{$>$\kern-.75em\lower1ex\hbox{$\sim$}}}}

\def\beq{\begin{equation}}
\def\eeq{\end{equation}}
\def\be{\begin{equation}}
\def\ee{\end{equation}}
\def\bea{\begin{eqnarray}}
\def\eea{\end{eqnarray}}

\def\to{\rightarrow}

\newcommand{\minigraph}[5][0.25in]{\begin{minipage}{#2}\begin{center}\includegraphics[width=#2]{#5}\\\vspace{#3}\hspace{#1}{\footnotesize #4}\end{center}\end{minipage}}

\title{Heavy Higgs Bosons at the LHC Upgrade}

\author{Tong Li}

\emailAdd{litong@nankai.edu.cn}
\affiliation{
School of Physics, Nankai University, Tianjin 300071, China}

\abstract{
We evaluate the discovery potential for the heavy Higgs bosons at the LHC energy upgrade with $\sqrt{s}=27$ TeV. We take degenerate mass spectrum and assume near the alignment limit in the Type-II Two Higgs Doublet Model for illustration.
We explore the observability of the heavy neutral Higgs bosons by examining the leading decay channel $H^0/A^0\to \tau^+\tau^-$ and the clean signals from $H^0\to W^+W^-, ZZ$ via gluon-gluon fusion production. The associated production of a top quark and a charged Higgs boson via $gb\to t H^\pm$ is adopted to predict the discovery potential of heavy charged Higgses. We also emphasize the potential importance of the electroweak production of Higgs boson pairs, i.e. $pp\to W^\ast \to H^\pm A^0$ and $pp\to Z^\ast/\gamma^\ast \to H^+ H^-$. They are only governed by pure electroweak gauge couplings and can provide complementary information to the conventional signals in the determination of the nature of the Higgs sector.
}

\begin{document}

\maketitle
\flushbottom
\newpage

\section{Introduction}
Since the milestone discovery of the Higgs boson at the CERN Large Hadron Collider (LHC)~\cite{Aad:2012tfa,Chatrchyan:2012xdj}, much attention has been drawn to the searches for new physics beyond the Standard Model (SM). Most of theoretical model constructions beyond the SM contain the extended Higgs sector, most notably in the minimal Supersymmetric Standard Model (MSSM)~\cite{Gunion:1984yn} and the composite Higgs model such as the little Higgs theory~\cite{ArkaniHamed:2002qy}. It is therefore strongly motivated to search for the new heavy Higgs bosons beyond the SM. Such efforts have been actively carried out, in particular in the LHC experiments.

While the LHC and its luminosity upgrade (HL-LHC) will continue the journey on searching for new physics in the next two decades, future higher energy hadron colliders, such as the energy upgrade for the LHC to 27 TeV C.M.~energy (HE-LHC)~\cite{Zimmermann:2018wdi,CidVidal:2018eel,Cepeda:2019klc} and the future circular collider of about 100 TeV C.M.~energy (FCC-hh)~\cite{Benedikt:2018csr}, are proposed to perform the direct searches at the energy frontier. In this paper, we set out an initial study for the discovery potential for the new heavy Higgs bosons at the HE-LHC.
We take the Type-II Two Higgs Doublet Model (2HDM) for illustration.

The leading search channel for the non-SM neutral Higgses comes from their single production, followed by their conventional
decays into pairs of SM particles. We thus study the gluon fusion processes $gg\to\phi\to \tau^+\tau^-, W^+W^-, ZZ$ and investigate the implication on the parameter space of the Type-II 2HDM model. For the charged Higgs heavier than top quark, the typical search channel is the associated production of a charged Higgs boson and top quark. The decay mode $H^\pm \to tb$ may suffer from large SM backgrounds but is dominant over other decays $H^\pm\to \tau^\pm\nu$ and $cs$, once kinematically accessible. For the sub-dominant decay $H^\pm\to \tau^\pm \nu$, the relevant SM backgrounds involve processes with $W^\pm\to \tau^\pm\nu$. The difference between the Yukawa coupling for $H^\pm$ and the gauge interaction for $W^\pm$, in terms of the spin correlation in tau decay, can be used to distinguish the signal from the SM backgrounds.

Although the above conventional signals for searching Higgs bosons are benefitted from large QCD production cross sections and simple kinematics, they all have a substantial dependence on additional 2HDM parameters, such as $\tan\beta$ and $\cos(\beta-\alpha)$. It is worth to emphasize the potential importance of the electroweak production of Higgs boson pairs, e.g. $pp\to W^\ast \to H^\pm A^0$ and $pp\to Z^\ast/\gamma^\ast \to H^+ H^-$. Their production cross sections are only governed by pure electroweak gauge couplings and quite complementary to the conventional signals in the determination of the Higgs nature.

The rest of the paper is organized as follows. In Sec.~\ref{sec:model}, we give a brief overview of the 2HDM and discuss the constraints on the parameters relevant for our study. In Sec.~\ref{sec:singleHA}, we analyze the single production of neutral Higgs bosons via gluon-gluon fusion and give the implication on the parameters of the Type-II 2HDM model. The prospect of probing single charged Higgs production is presented in Sec.~\ref{sec:singleHpm}. In Sec.~\ref{sec:pairHiggs}, we study the signatures of non-SM Higgses pair production through pure electroweak interactions. Finally, in
Sec.~\ref{sec:Concl} we summarize our main results.

\section{Two Higgs Doublet Model}
\label{sec:model}

Two Higgs Doublet Model~\cite{Branco:2011iw} is a good representative prototype to study the Higgs boson properties beyond the SM.
In the 2HDM, the Higgs sector is composed of two SU$(2)_L$ scalar doublets
\begin{eqnarray}
H_i = \left(
        \begin{array}{c}
          h_i^+ \\
          (v_i + h_i + i P_i)/\sqrt{2} \\
        \end{array}
      \right), \ \ i=1,2.
\end{eqnarray}
After the electroweak symmetry breaking (EWSB), there are four more Higgs bosons ($H^0, A^0, H^\pm$) besides the SM-like Higgs boson ($h^0$) in the particle spectrum
\begin{eqnarray}
&&\left(
  \begin{array}{c}
    H^0 \\
    h^0 \\
  \end{array}
\right)
=
\left(
  \begin{array}{cc}
    \cos\alpha & \sin\alpha \\
    -\sin\alpha & \cos\alpha \\
  \end{array}
\right)
\left(
  \begin{array}{c}
    h_1 \\
    h_2 \\
  \end{array}
\right), \nonumber \\
&& A^0 = -\sin\beta P_1 + \cos\beta P_2, \ \ H^\pm = -\sin\beta h_1^\pm + \cos\beta h_2^\pm .
\end{eqnarray}
Here, the important parameter is defined as $\tan\beta=v_2/v_1$ with $\sqrt{v_1^2+v_2^2}=v=246$ GeV.
Because of the absence of new physics signals from the searches at the LHC, we demand that the non-SM Higgses are all heavier than $h^0$ and take their masses as free parameters.
Certain discrete symmetries between the two doublets are often imposed to avoid unwanted flavor-changing-neutral currents (FCNC).

Motivated by the construction of the minimal Supersymmetric Standard Model (MSSM), we assume the Type-II 2HDM in which $H_1$ only couples to the down-type quarks and leptons and $H_2$ only couples to the up-type quarks. Their couplings to the SM fermions behave as
\begin{eqnarray}
&&g_{H^0u\bar{u}}={\sin\alpha\over \sin\beta}=\cos(\beta-\alpha)-\cot\beta\sin(\beta-\alpha), \nonumber  \\
&&g_{H^0d\bar{d}}=g_{H^0l\bar{l}}={\cos\alpha\over \cos\beta}=\cos(\beta-\alpha)+\tan\beta\sin(\beta-\alpha); \nonumber \\
&&g_{A^0u\bar{u}}=-i\cot\beta\gamma_5, \ \ \ g_{A^0d\bar{d}}=g_{A^0l\bar{l}}=-i\tan\beta\gamma_5; \nonumber \\
&&g_{H^+\bar{u}d}=-{i\over \sqrt{2}v}V_{ud}^\ast\left[m_d\tan\beta(1+\gamma_5)+m_u\cot\beta(1-\gamma_5)\right], \nonumber \\
&&g_{H^-u\bar{d}}=-{i\over \sqrt{2}v}V_{ud}\left[m_d\tan\beta(1-\gamma_5)+m_u\cot\beta(1+\gamma_5)\right],\nonumber \\
&&g_{H^+\bar{\nu}l}=-{i\over \sqrt{2}v}m_l\tan\beta(1+\gamma_5), \ \ \ g_{H^-\nu\bar{l}}=-{i\over \sqrt{2}v}m_l\tan\beta(1-\gamma_5),
\end{eqnarray}
with a normalization factor $im_{u,d,l}/v$ for neutral Higgses. The couplings between neutral Higgses and two gauge bosons are $g_{H^0VV}=\cos(\beta-\alpha)$ and $g_{A^0VV}=0$. As such, the parameters involved in our analyses include $\tan\beta$, $\cos(\beta-\alpha)$, and the relevant Higgs masses under consideration.

As intimated before, we identify the lighter CP-even scalar $h^0$ as the SM-like Higgs observed at the LHC. This, together with the absence of exotic decays of the 125 GeV Higgs boson, implies the alignment limit~\cite{Carena:2013ooa,Bernon:2015qea}. We will take the alignment limit $\cos(\beta-\alpha)=0$ or assume the value of $\cos(\beta-\alpha)$ not far away from the alignment in the following analysis.
The theoretical consideration of vacuum stability~\cite{Gunion:2002zf} and unitarity~\cite{Ginzburg:2005dt} and the measurement of electroweak precision observables~\cite{Haller:2018nnx} suggest small mass splittings among the four non-SM Higgses. We thus assume degenerate heavy Higgs mass spectrum (unless otherwise stated) and forbid exotic Higgs decay modes~\cite{Kling:2016opi,Kling:2018xud,Khachatryan:2016are,Aaboud:2017cxo,Aaboud:2018eoy}.

In addition, the non-SM Higgs sector is strongly constrained by various flavor physics measurements such as the $b\to s$ transitions~\cite{Amhis:2016xyh}. The charged Higgs boson in Type-II 2HDM is in particular required to be heavier than about 600 GeV~\cite{Arbey:2017gmh}. This constraint can be relaxed by the cancellation between the charged Higgs contribution and new contributions to the flavor observables from other sectors in new physics models~\cite{Chen:2009cw,Han:2013mga}. As we focus on the collider search of heavy Higgs bosons in this paper, we will not pursue the flavor constraints explicitly. We will individually take into account LHC constraints for the specific decay channels of heavy Higgses we consider in the following.

\section{Single Neutral Higgs Production}
\label{sec:singleHA}

\begin{figure}[t]
\begin{center}
\minigraph{7cm}{-0.05in}{(a)}{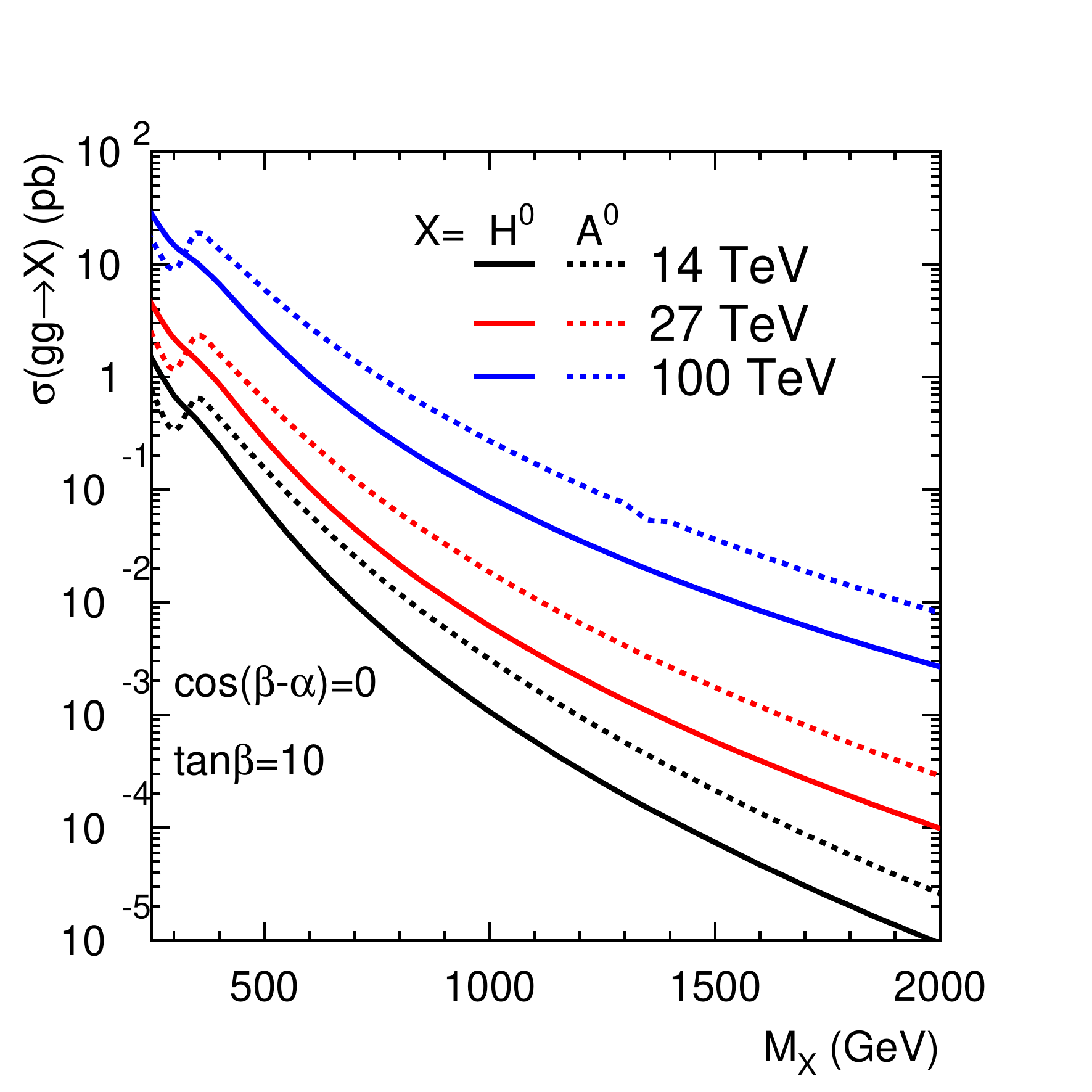}
\minigraph{7cm}{-0.05in}{(b)}{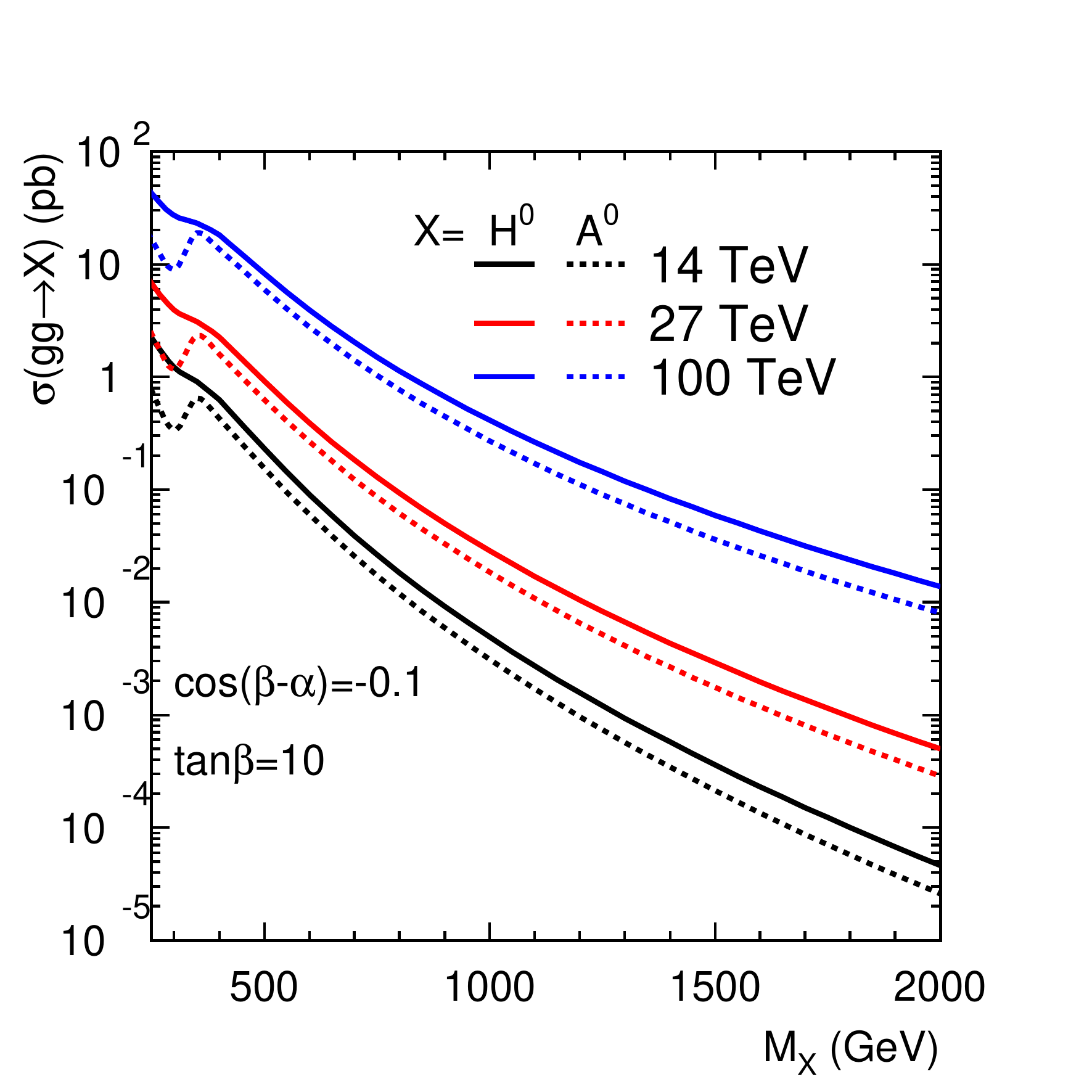}\\
\minigraph{7cm}{-0.05in}{(c)}{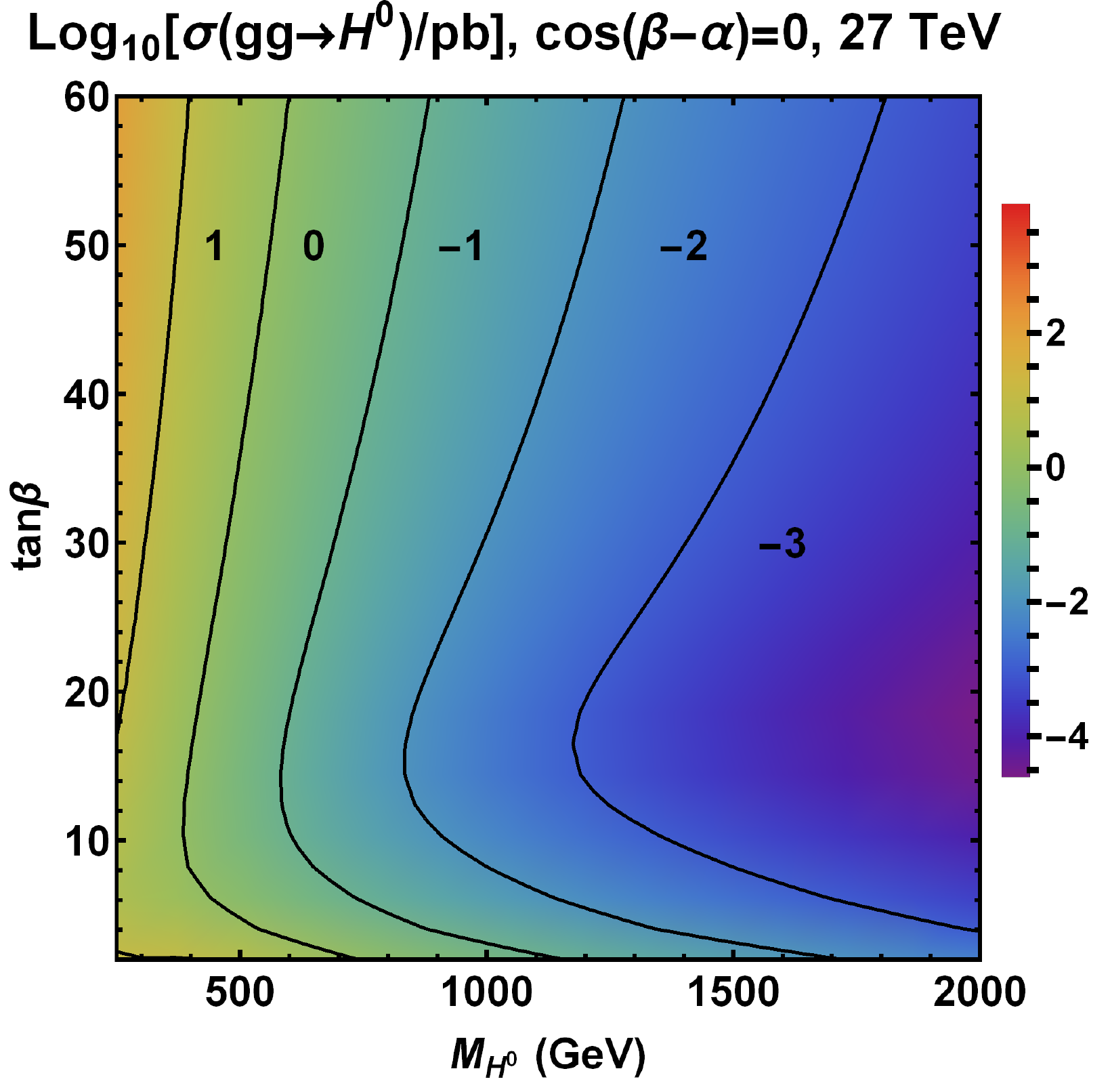}
\minigraph{7cm}{-0.05in}{(d)}{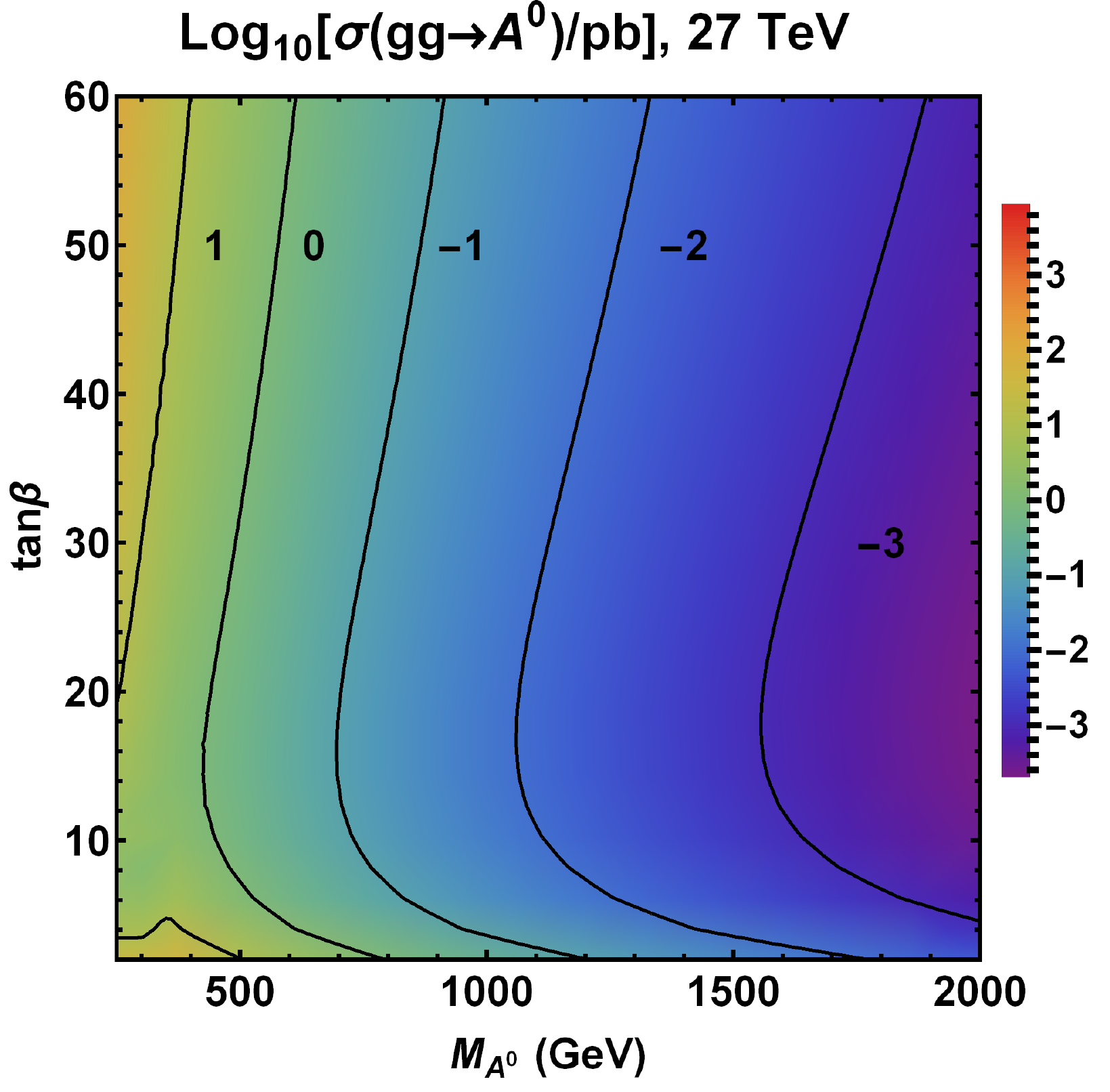}
\end{center}
\caption{Top: Total production cross section versus the Higgs boson mass for $gg \to H^0, A^0$ with $\cos(\beta-\alpha)=0$ (a) or $\cos(\beta-\alpha)=-0.1$ (b) and $\tan\beta=10$ at $pp$ collider with 14 TeV, 27 TeV and 100 TeV. Bottom: The cross section indicated by contour lines in the plane of $\tan\beta$ versus the Higgs boson mass for $gg\to H^0$ with $\cos(\beta-\alpha)=0$ (c) and $gg\to A^0$ (d) at the 27 TeV LHC.
}
\label{fig:xsecggHA}
\end{figure}

Just like the Higgs boson discovery, the leading production channel for a heavy neutral Higgs boson is through the gluon fusion
\beq
gg\to H^0, \ A^0.
\eeq
These channels are benefitted from the large gluon luminosity at higher energies and the favorable phase space for a single particle production.
We show the production cross sections versus Higgs mass (from 250 GeV to 2 TeV) at the 14 TeV LHC, 27 TeV LHC, as well as the 100 TeV collider in Fig.~\ref{fig:xsecggHA}. The cross sections are obtained at NNLO in QCD using default SusHi~\cite{Harlander:2012pb} and LHAPDF~\cite{Buckley:2014ana} with the alignment limit $\cos(\beta-\alpha)=0$ or $\cos(\beta-\alpha)=-0.1$ (note that the $gg\to A^0$ production does not depend on $\cos(\beta-\alpha)$). We see that the total production cross section at 27 TeV LHC ranges from 4 (2.8) pb at $M_{H^0(A^0)}=250$ GeV to $1 \ (3) \times 10^{-4}$ pb at $M_{H^0(A^0)}=2$ TeV for $\tan\beta=10$ in the alignment limit. It increases by four times at $M_{H^0/A^0}=500$ GeV and by eight times at $M_{H^0/A^0}=1.5$ TeV from 14 TeV to 27 TeV C.M. energy.

We explore the observability of the heavy neutral Higgs bosons by examining the specific decay channels.

\subsection{$ H^0, A^0 \to \tau^+ \tau^-$}
\label{sec:HAtautau}

We first consider the decay $H^0,A^0\to \tau^+\tau^-$, followed by the $\tau$'s leading 2-body decay mode $\tau^\pm\to \pi^\pm \nu_\tau$ with the branching fraction being ${\rm BR}(\tau^\pm\to \pi^\pm \nu_\tau)=0.11$. The $\tau$-spin correlation is maximized in this decay channel. Our signal thus consists of two opposite-sign pions and missing neutrinos. The irreducible SM backgrounds are from diboson productions $W^+W^-,ZZ\to \tau^+\tau^-\nu\bar{\nu}$ and the reducible contribution is $W^\pm Z\to \tau^+\tau^-\ell^\pm \nu_\ell$ with the additional charged lepton $\ell^\pm$ vetoed if $p_T(\ell)>7 \ {\rm GeV}, |\eta(\ell)|<3.5$.
We use MadGraph5\_aMC@NLO~\cite{Alwall:2014hca} to generate signal and backgrounds events, and TAUOLA~\cite{Jadach:1993hs} interfaced with Pythia~\cite{Sjostrand:2006za} to simulate tau lepton decay.

\begin{figure}[h!]
\begin{center}
\minigraph{7cm}{-0.05in}{(a)}{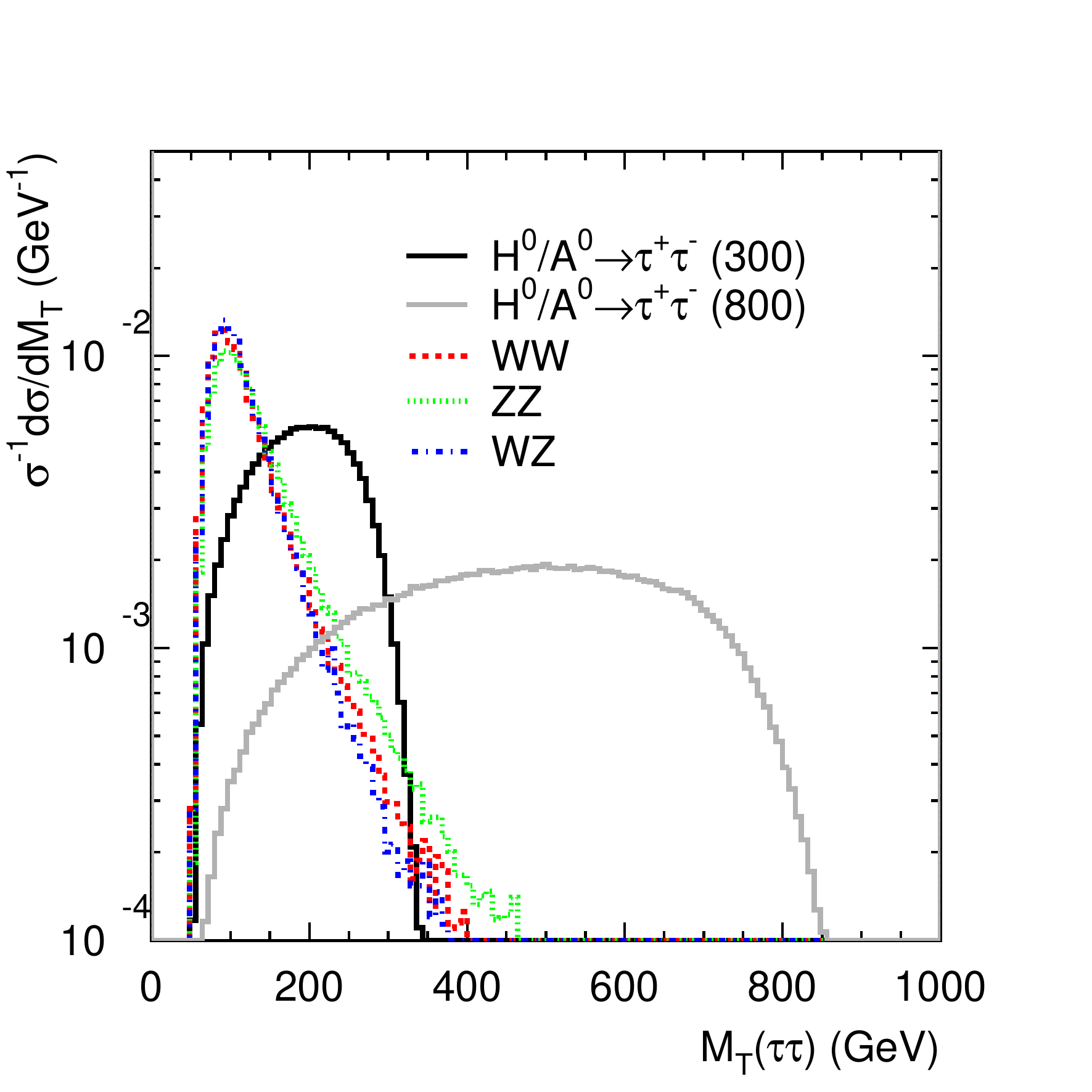}
\minigraph{7cm}{-0.05in}{(b)}{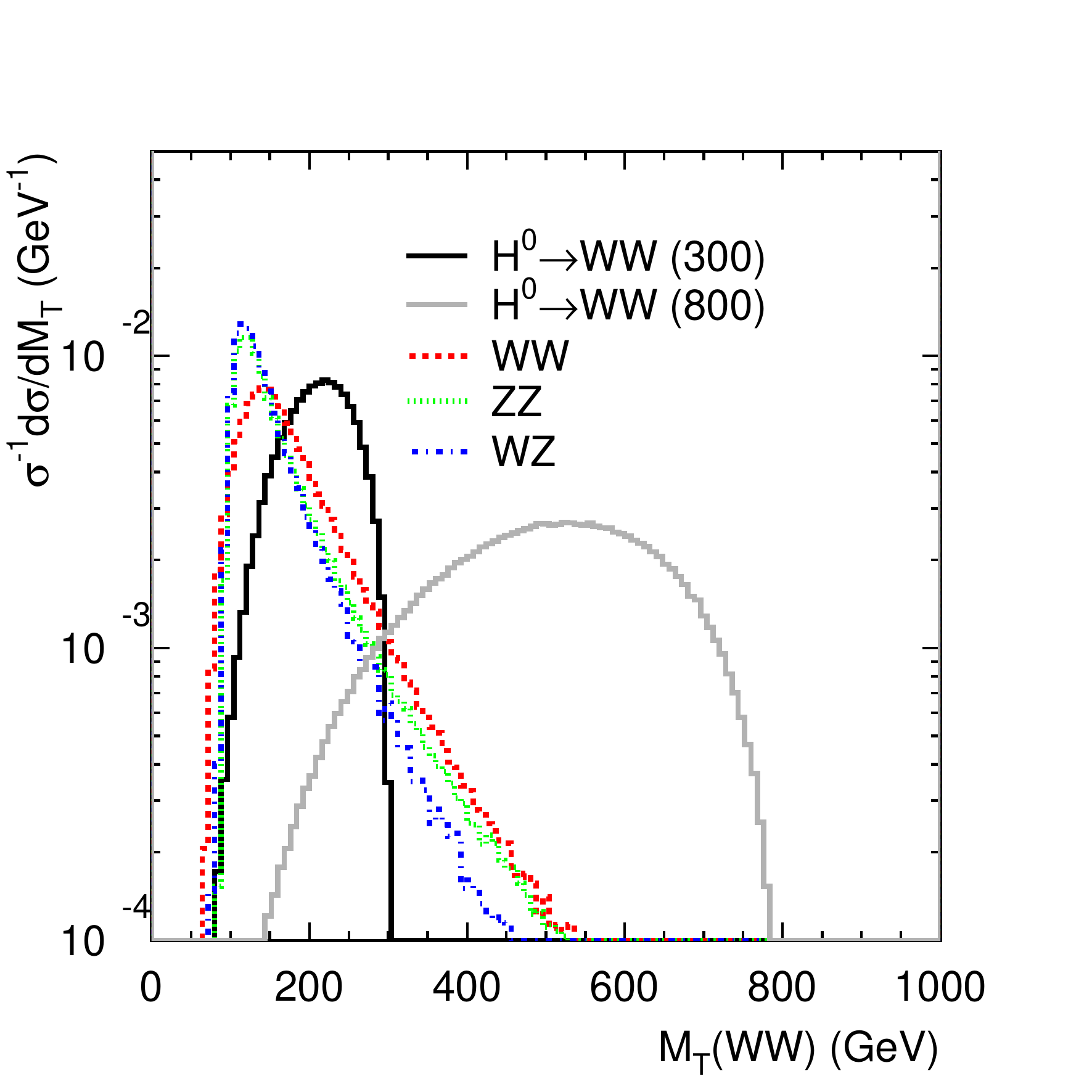}
\end{center}
\caption{The differential cross section distributions of the total transverse mass $M_T(\tau^+\tau^-)$ for the signal $gg\to H^0/A^0\to \tau^+\tau^-$ (a) and the $WW$ transverse mass $M_T(WW)$ for the signal $gg\to H^0\to W^+W^-$ (b), together with SM backgrounds at the 27 TeV LHC. }
\label{ggha-MTdis}
\end{figure}

We follow the search strategy recently carried out by the ATLAS collaboration~\cite{Aaboud:2017sjh}, and adopt the acceptance cuts as
\begin{eqnarray}
p_T(\pi)\geq 25 \ {\rm GeV}; \ \ \ |\eta(\pi)|<2.5; \ \ \ \Delta R_{\pi\pi}\geq 0.4.
\end{eqnarray}
The mass of the heavy Higgs resonance can be read from the edge of the total transverse mass
\begin{eqnarray}
M_T(\tau^+\tau^-)=\sqrt{(p_T(\pi_1)+p_T(\pi_2)+\cancel{E}_T)^2-(\vec{p}_T(\pi_1)+\vec{p}_T(\pi_2)+\vec{\cancel{p}}_T)^2},
\end{eqnarray}
as shown in Fig.~\ref{ggha-MTdis} (a).
To enhance the acceptance of our signal, we further require the events pass the following selection cuts, namely
\begin{itemize}
\item missing energy cut: $\cancel{E}_T>40 \ {\rm GeV}$,
\item minimal $p_T$ cut on the two charged pions: $p_T^{\rm min}(\pi)>65 \ {\rm GeV}$,
\item azimuthal angle cut for the back-to-back pions in the transverse plane: \\
$|\Delta\phi(\vec{p}_{T\pi_1},\vec{p}_{T\pi_2})|>2.7$.
\end{itemize}
The cut efficiencies for the signal and SM backgrounds are collected in Table~\ref{cuteff-ggha-tata}.
The dominant backgrounds after $\cancel{E}_T, p_T$ cuts are the irreducible backgrounds $WW, ZZ$. The $ZZ$ background with
one $Z$ decaying invisibly and the other decaying to two tau leptons can be further suppressed by the azimuthal angle cut.

\begin{table}[tb]
\begin{center}
\begin{tabular}{|c|c|c|c|c|c|}
\hline
cut efficiencies  & basic cuts & $\cancel{E}_T$ & $p_T^\pi$  & $\Delta\phi$
\\ \hline
$H^0/A^0\to \tau^+\tau^- (300)$ & 0.44 & 0.19 & 0.034 & 0.034 \\ 
$H^0/A^0\to \tau^+\tau^- (800)$ & 0.72 & 0.54 & 0.39 & 0.39 \\ 
\hline \hline
$WW$ & 0.024 & 0.0055 & 0.00068 & 0.00063 \\ 
\hline
$ZZ$ & 0.084 & 0.044 & 0.0019 & negligible \\ 
\hline
$WZ$ & 0.0094 & 0.0037 & $9\times 10^{-5}$ & negligible\\ 
\hline
\end{tabular}
\end{center}
\caption{The cut efficiencies for $gg\to H^0/A^0\to \tau^+\tau^-$ and the SM backgrounds after consecutive cuts with $\tau^\pm\to \pi^\pm \nu_\tau$ channel at the 27 TeV LHC. We take $M_{H^0}=M_{A^0}=300$ or 800 GeV.}
\label{cuteff-ggha-tata}
\end{table}

\begin{figure}[h!]
\begin{center}
\minigraph{7cm}{-0.05in}{(a)}{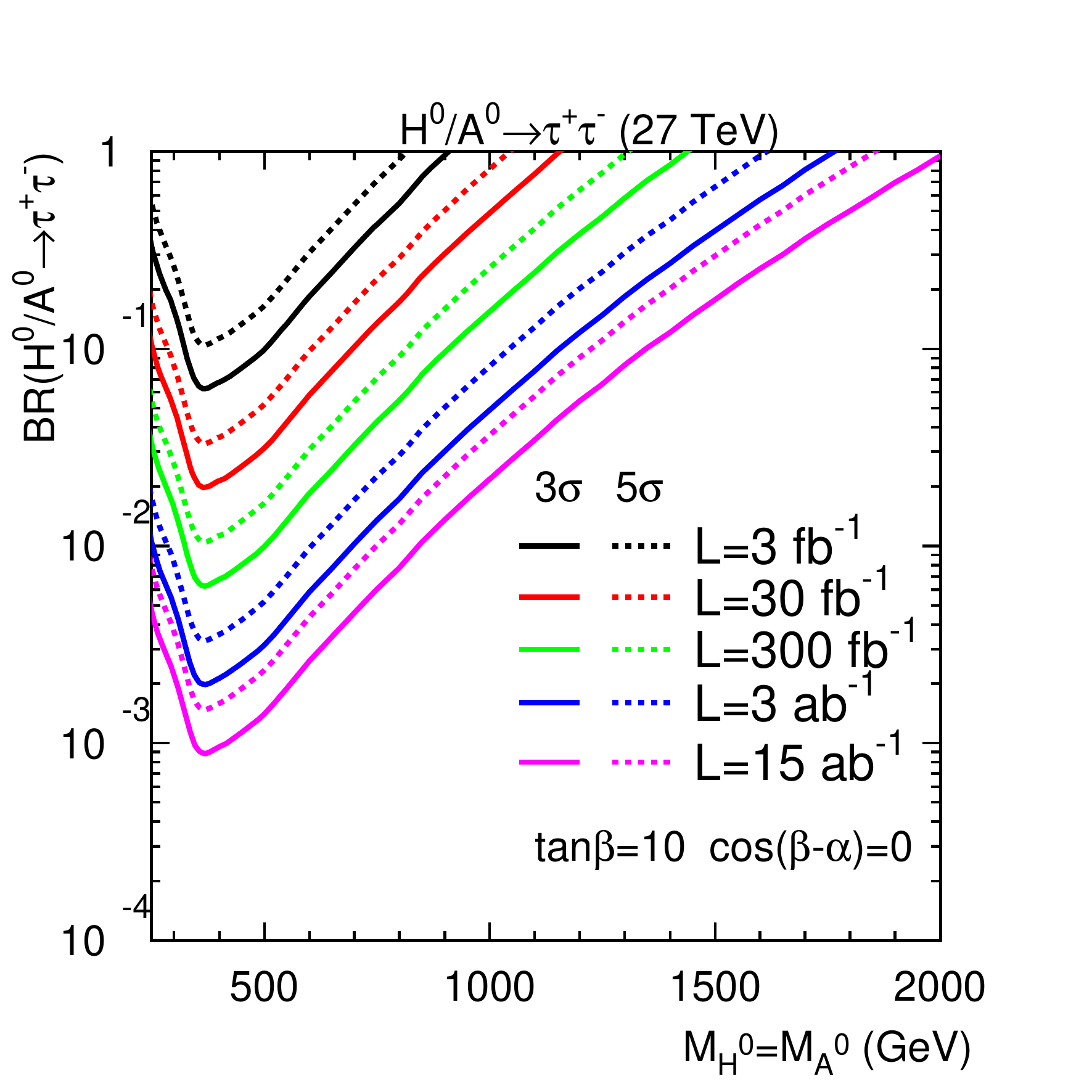}
\minigraph{6.2cm}{-0.05in}{(b)}{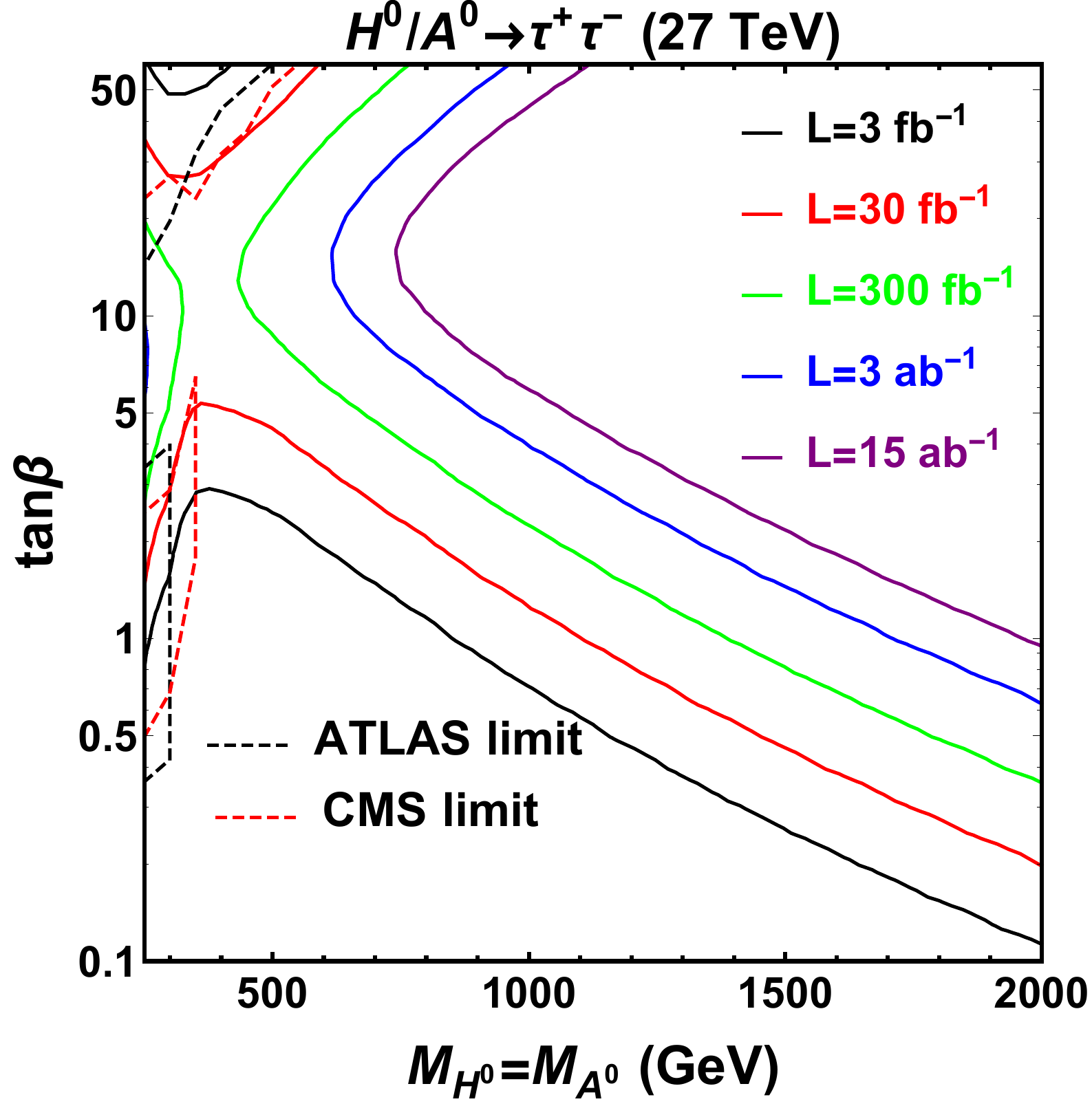}
\end{center}
\caption{
Left: Reach of BR$(H^0/A^0\to \tau^+\tau^-)$ as a function of $M_{H^0/A^0}$ for $gg\to H^0/A^0\to \tau^+\tau^-$ channel at the 27 TeV LHC. We assume $\tan\beta=10$ and $\cos(\beta-\alpha)=0$.
Right: Discovery contour in $\tan\beta$ versus $M_{H^0/A^0}$ plane for $gg\to H^0/A^0\to \tau^+\tau^-$ in the alignment limit, with realistic BR$(H^0/A^0\to \tau^+\tau^-)$ under the assumption of $M_{H^0}=M_{A^0}=M_{H^\pm}$.
The excluded regions in the Type-II 2HDM are indicated by the dashed curves, based on $gg\to H^0/A^0\to \tau\tau$ search at the 13 TeV LHC~\cite{Aaboud:2017sjh,Sirunyan:2018zut}.
}
\label{sigma-ggha-tata}
\end{figure}

Next we show the prospect of probing $gg\to H^0/A^0\to \tau^+\tau^-$ channel at the 27 TeV LHC in the context of Type-II 2HDM.
The left panel of Fig.~\ref{sigma-ggha-tata} displays the reachable limit of BR$(H^0/A^0\to \tau^+\tau^-)$ as a function of $M_{H^0/A^0}$ with $\tan\beta=10$ and $\cos(\beta-\alpha)=0$. The solid and dashed curves correspond to 3$\sigma$ significance and 5$\sigma$ discovery, respectively. With 15 ab$^{-1}$ luminosity, the branching fraction limit of $H^0/A^0\to \tau^+\tau^-$ can be reached as low as $1.5\times 10^{-3}$ for $M_{H^0/A^0}\simeq 350$ GeV and $H^0, A^0$ with the mass of about 1.85 TeV can be probed for 5$\sigma$ discovery if BR$(H^0/A^0\to \tau^+\tau^-)=1$. As the decays of $H^0, A^0$ into heavy quarks are dominant for small and moderate $\tan\beta$ if kinematically accessible~\cite{Hajer:2015gka,Craig:2016ygr} and the decay into $b\bar{b}$ dominates over $\tau\tau$ mode for large $\tan\beta$, the realistic branching fraction of $H^0/A^0\to \tau^+\tau^-$ cannot reach the order of unity. We use package 2HDMC~\cite{Eriksson:2009ws} to calculate all 2HDM branching fractions below.

The LHC provided the observed 95\% CL upper limit on the gluon-gluon fusion production cross section times the branching fraction of a scalar boson decay into $\tau\tau$ at $\sqrt{s}=13$ TeV, corresponding to an integrated luminosity of about 36 fb$^{-1}$~\cite{Aaboud:2017sjh,Sirunyan:2018zut}. We recast the limit in the plane of $\tan\beta$ versus $M_{H^0/A^0}$ for the Type-II 2HDM as shown by dashed curves in Fig.~\ref{sigma-ggha-tata} (b). The current exclusion is about $M_{H^0/A^0} \simeq 300$ GeV for $\tan\beta=1$, and $M_{H^0/A^0} \simeq 500$ GeV for $\tan\beta=50$.
With realistic BR$(H^0/A^0\to \tau^+\tau^-)$ under the assumption of $M_{H^0}=M_{A^0}=M_{H^\pm}$, the discovery region of 27 TeV LHC in the alignment limit is displayed in Fig.~\ref{sigma-ggha-tata} (b) for $gg\to H^0/A^0\to \tau^+\tau^-$. The regions to the left of the curves can be covered by 5$\sigma$ discovery, corresponding to different luminosities. One can see that the reach at 27 TeV LHC can cover most of the region with small $\tan\beta$. The wedge region with $\tan\beta\sim 10$ loses sensitivity due to the suppression of the production cross section. For $\tan\beta=1 \ (10) \ [50]$, the 27 TeV LHC can probe the neutral Higgs as heavy as 2 TeV (800 GeV) [1.1 TeV] with the luminosity of 15 ab$^{-1}$.

\subsection{$ H^0 \to W^+W^-, ZZ$}
\label{sec:HVV}

By far, the cleanest signals for heavy new physics would be the leptonic final states from the $W/Z$ decays. We now utilize those channels to search for the CP-even Higgs $H^0$. The basic requirements for the leptons are
\begin{eqnarray}
&&p_T(\ell)\geq 30 \ {\rm GeV}, \ \ \  |\eta(\ell)|<2.5, \ \ \ \Delta R_{\ell\ell}\geq 0.4,
\end{eqnarray}
and we select the events satisfying
\begin{eqnarray}
&&\cancel{E}_T>40 \ {\rm GeV}, \ \ \ p_T^{\rm min}(\ell)>65 \ {\rm GeV}, \ \ \ M_{\ell\ell}>M_{H^0}/3, 
\end{eqnarray}
for $H^0\to W^+W^-$ channel. The mass of $H^0$ resonance in $WW$ channel can be reconstructed by the $WW$ transverse mass
\begin{eqnarray}
M_T(W^+W^-)=\sqrt{(E_T^{\ell\ell}+\cancel{E}_T)^2-(\vec{p}_T(\ell_1)+\vec{p}_T(\ell_2)+\vec{\cancel{p}}_T)^2}, \ E_T^{\ell\ell}=\sqrt{|\vec{p}_T^{\ell\ell}|^2+m_{\ell\ell}^2}, \nonumber \\
\end{eqnarray}
as shown in Fig.~\ref{ggha-MTdis} (b).
The SM backgrounds are the same as those for $\tau^+\tau^-$ channel but with gauge bosons' leptonic decay to electron/muon.
The $ZZ$ background has the opposite-sign lepton pairs $\ell^+\ell^-$ from $Z$ boson decay and can be further reduced by vetoing the invariant mass of opposite sign leptons if $|M_{\ell\ell}-M_Z|<10$ GeV. For $H^0\to ZZ$ channel, we simply require
\begin{eqnarray}
p_T^{\rm min}(\ell)>50 \ {\rm GeV}, \ \ \ |M_{4\ell}-M_{H^0}|<M_{H^0}/10,
\end{eqnarray}
for the minimal lepton $p_T$ and the invariant mass of the four leptons.
The cut efficiencies are given in Tables~\ref{cuteff-ggh-ww} and \ref{cuteff-ggh-zz} for $WW$ and $ZZ$ channels, respectively. One can see that the $Z$ boson veto and the mass window requirement for $H^0$ resonance significantly suppress the $ZZ$ background for $H^0\to W^+W^-$ and $H^0\to ZZ$, respectively.

\begin{table}[tb]
\begin{center}
\begin{tabular}{|c|c|c|c|c|c|c|}
\hline
cut efficiencies  & basic cuts & $\cancel{E}_T$ & $p_T^\ell$  & $M_Z$ veto & $M_{\ell\ell}$
\\ \hline
$H^0\to W^+W^- (300)$ & 0.52 & 0.35 & 0.082 & 0.082 & 0.082   \\
$H^0\to W^+W^- (800)$ & 0.79 & 0.66 & 0.54 & 0.54 & 0.50   \\
\hline \hline
$WW$ (300) & 0.23 & 0.1 & 0.016 & 0.016 & 0.016  \\
$WW$ (800) & 0.23 & 0.1 & 0.016 & 0.016 & 0.0071  \\
\hline
$ZZ$ (300) & 0.33 & 0.18 & 0.015 & 0.00099 & 0.00072 \\
$ZZ$ (800) & 0.33 & 0.18 & 0.015 & 0.00099 & negligible \\
\hline
$WZ$ (300) & 0.046 & 0.02 & 0.0012 & 0.00048 & 0.00047 \\
$WZ$ (800) & 0.046 & 0.02 & 0.0012 & 0.00048 & 0.00021 \\
\hline
\end{tabular}
\end{center}
\caption{The cut efficiencies for $gg\to H^0\to W^+W^-$ and the SM backgrounds after consecutive cuts at the 27 TeV LHC. We take $M_{H^0}=300$ or 800 GeV.}
\label{cuteff-ggh-ww}
\end{table}

\begin{table}[tb]
\begin{center}
\begin{tabular}{|c|c|c|c|}
\hline
cut efficiencies  & basic cuts & $p_T^\ell$  & $M_{4\ell}$
\\ \hline
$H^0\to ZZ (300)$ & 0.3 & 0.053 & 0.053  \\
$H^0\to ZZ (800)$ & 0.69 & 0.58 & 0.58 \\
\hline \hline
$ZZ (300)$ & 0.12 & 0.0097 & 0.0014 \\
$ZZ (800)$ & 0.12 & 0.0097 & 0.00081\\
\hline
\end{tabular}
\end{center}
\caption{The cut efficiencies for $gg\to H^0\to ZZ$ and the SM backgrounds after consecutive cuts at the 27 TeV LHC. We take $M_{H^0}=300$ or 800 GeV.}
\label{cuteff-ggh-zz}
\end{table}

\begin{figure}[h!]
\begin{center}
\minigraph{7cm}{-0.05in}{(a)}{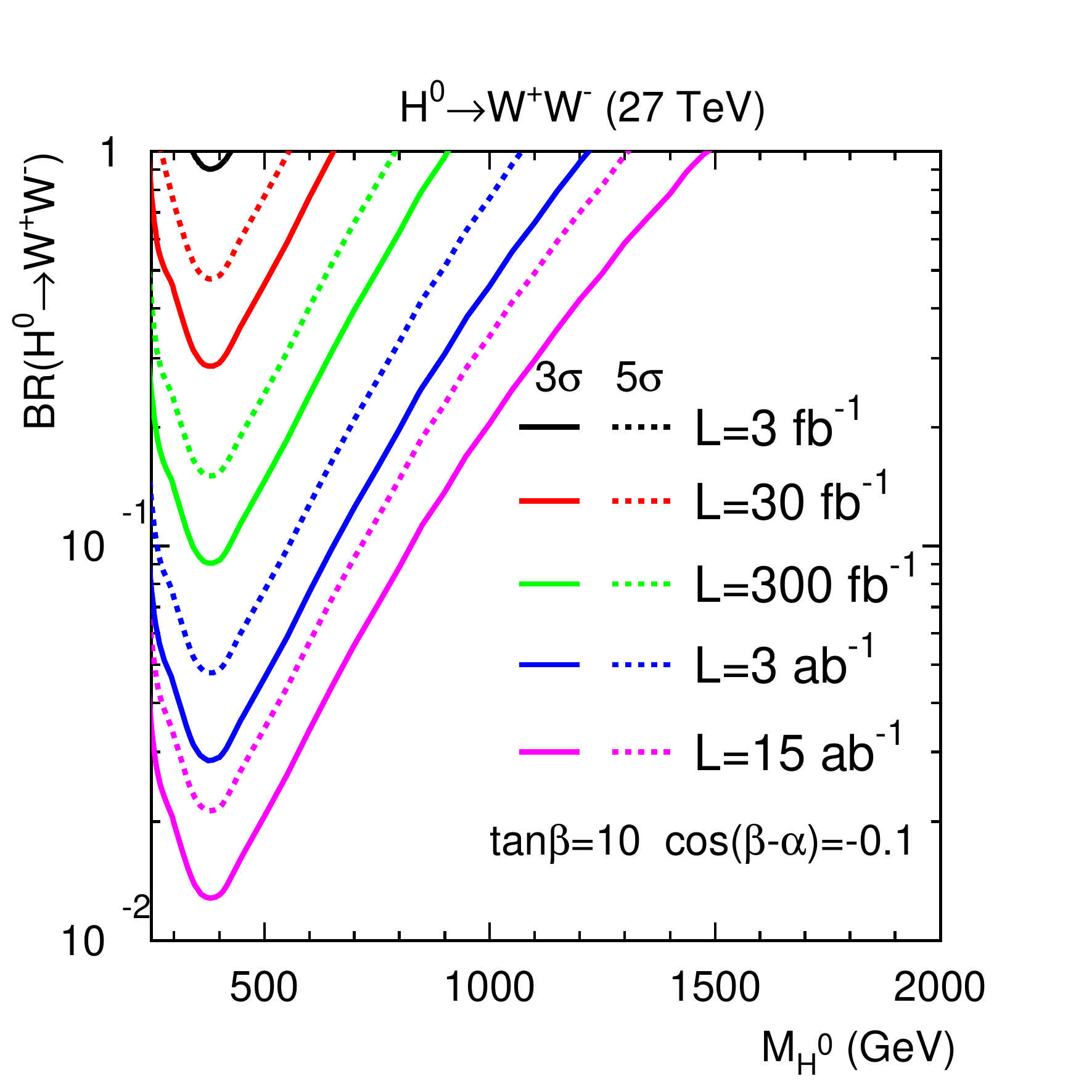}
\minigraph{6.2cm}{-0.05in}{(b)}{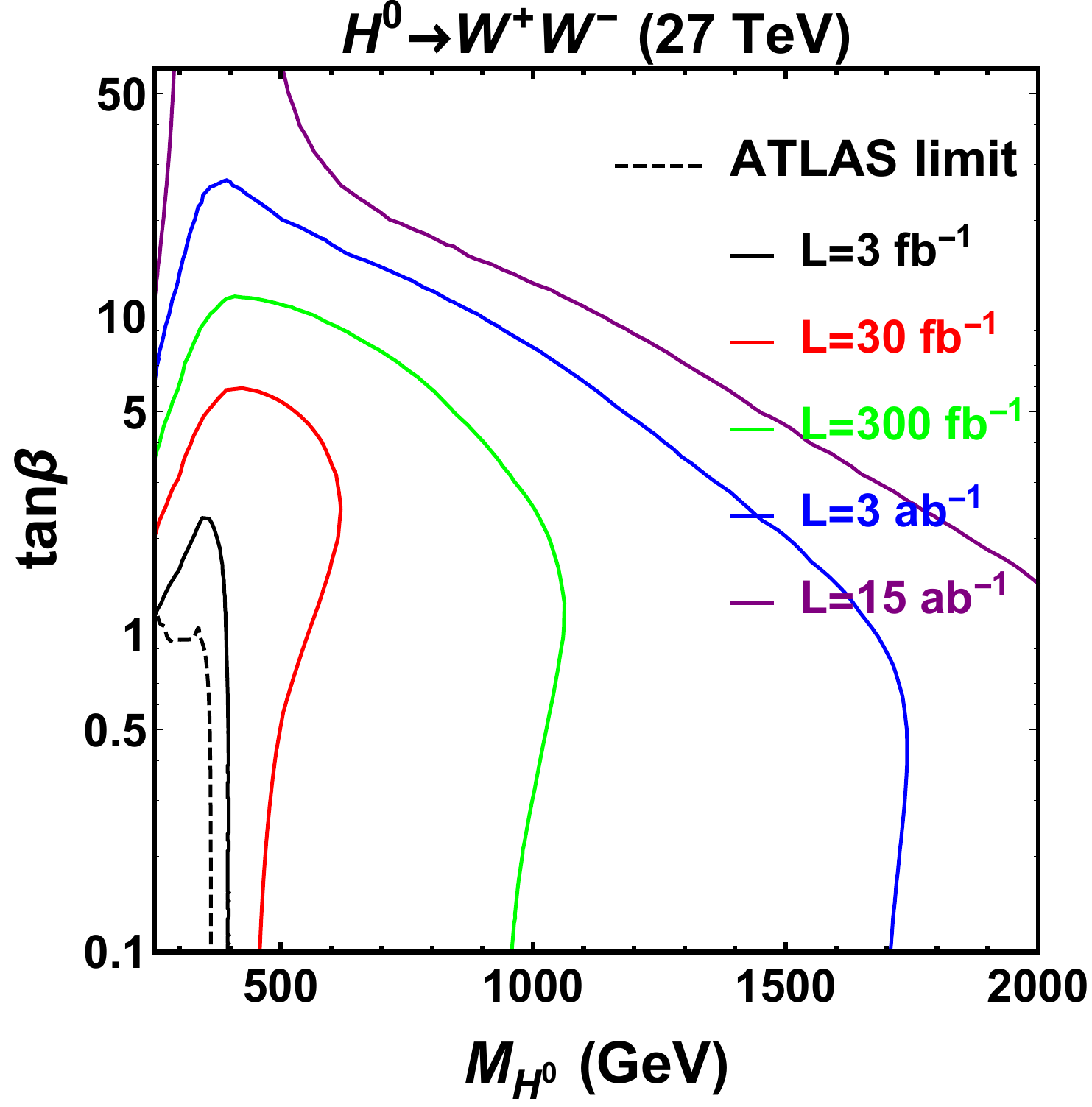}\\
\minigraph{7cm}{-0.05in}{(c)}{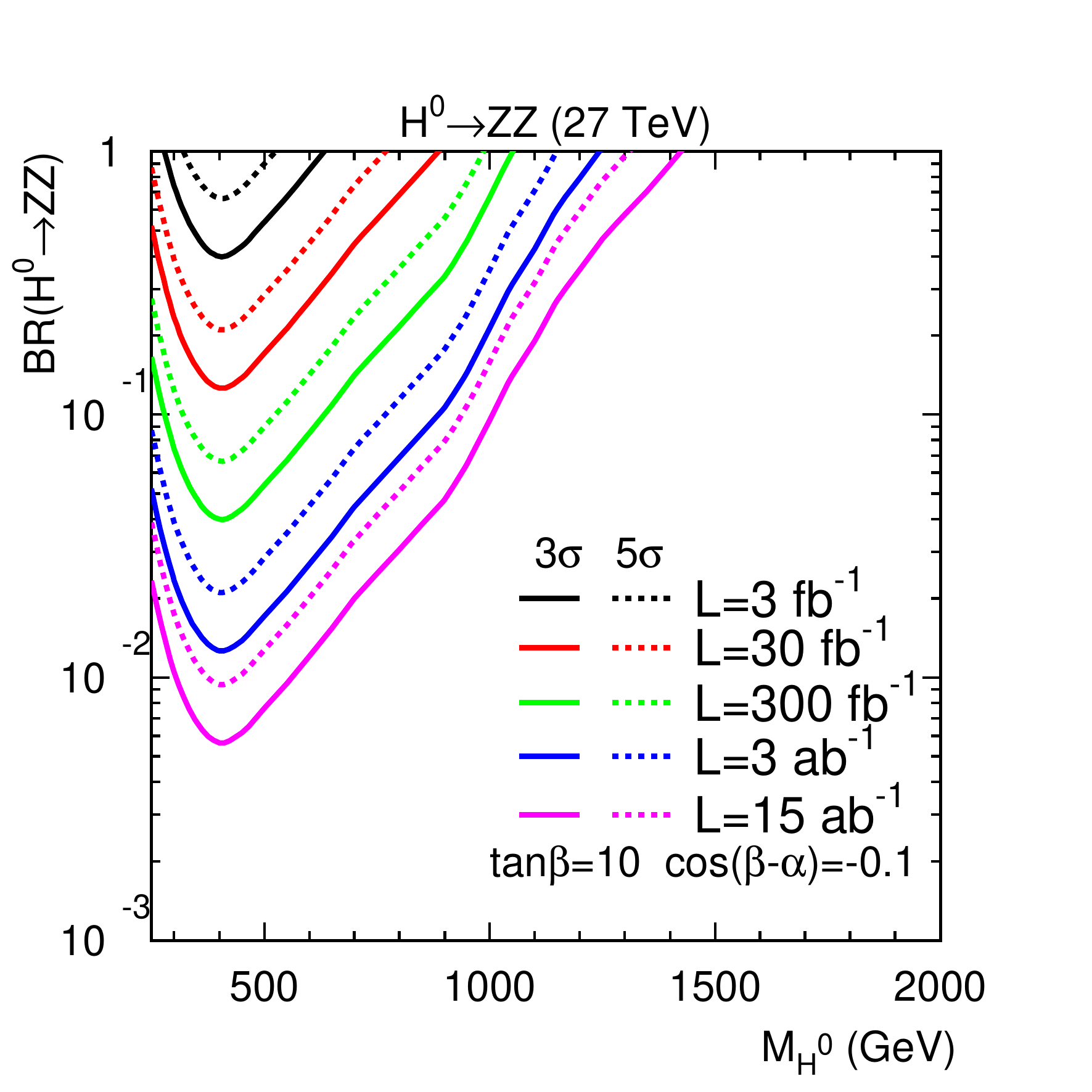}
\minigraph{6.2cm}{-0.05in}{(d)}{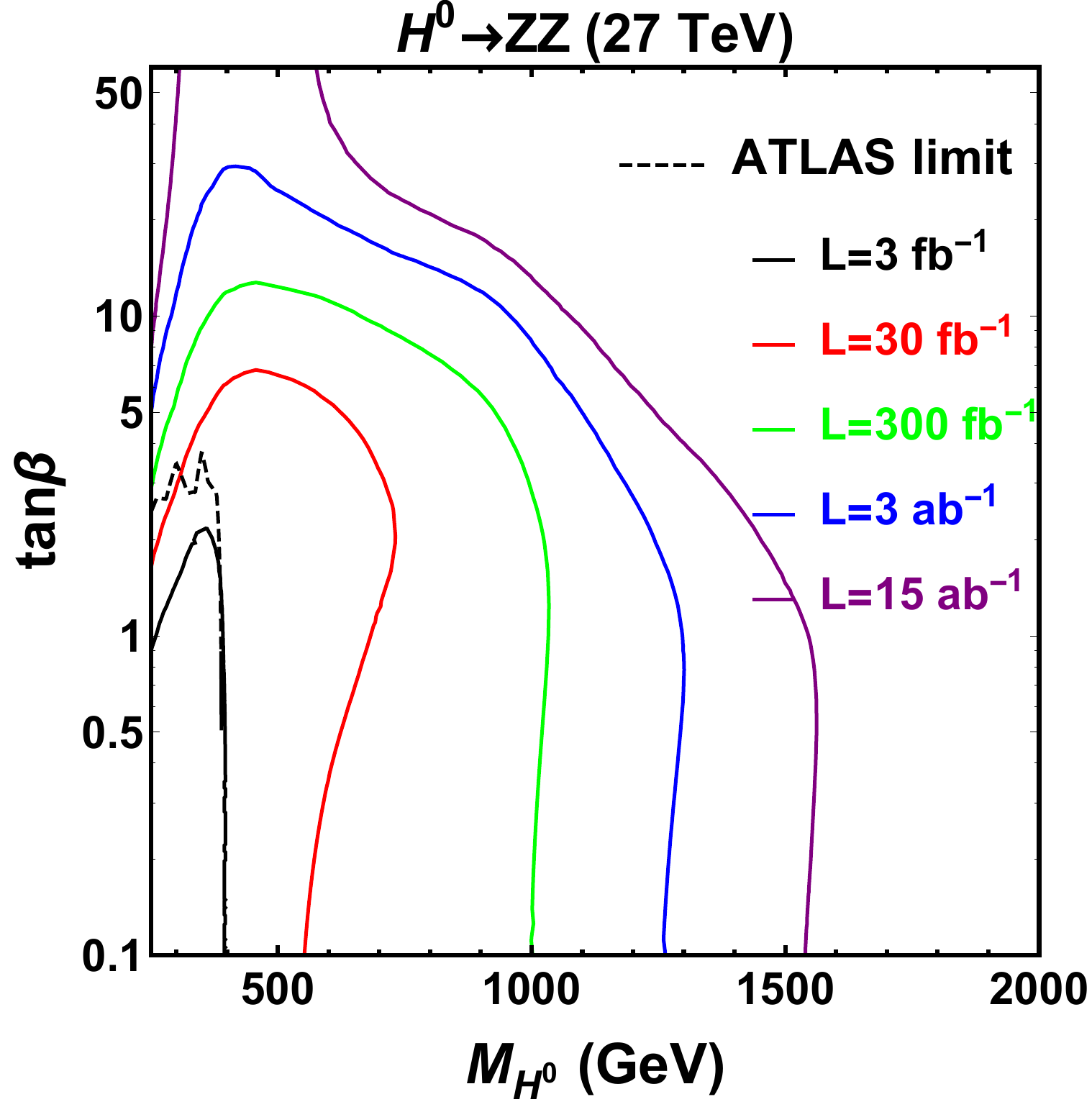}
\end{center}
\caption{
Left panels: Reach of BR$(H^0\to W^+W^-)$ (a) and BR$(H^0\to ZZ)$ (c) as a function of $M_{H^0}$ at the 27 TeV LHC. We assume $\tan\beta=10$ and $\cos(\beta-\alpha)=-0.1$.
Right panels: Discovery contour in $\tan\beta$ versus $M_{H^0}$ plane for $gg\to H^0\to W^+W^-$ (b) and $gg\to H^0\to ZZ$ (d), with realistic BR$(H^0\to W^+W^-/ZZ)$ under the assumption of $\cos(\beta-\alpha)=-0.1$.
The excluded regions in the Type-II 2HDM are indicated by the dashed curves, based on $gg\to H^0\to WW,ZZ$ search at the 13 TeV LHC~\cite{Aaboud:2017gsl,Aaboud:2017rel}.
}
\label{sigma-ggh-wz}
\end{figure}

The decays of $H^0\to W^+W^-, ZZ$ are present away from the alignment limit, and can dominate with larger values of $|\cos(\beta-\alpha)|$.
Assuming $\cos(\beta-\alpha)=-0.1$ and $\tan\beta=10$, in the left panels of Fig.~\ref{sigma-ggh-wz}, we show the reach of BR$(H^0\to W^+W^-, ZZ)$ as a function of $M_{H^0}$ at the 27 TeV LHC. The minimal branching fraction that can be reached with 15 ab$^{-1}$ luminosity is around $(1-2)\times 10^{-2}$.

The exclusion contours for $H^0$ decay to the SM gauge bosons by the 13 TeV LHC~\cite{Aaboud:2017gsl,Aaboud:2017rel} are added in the right panels of Fig.~\ref{sigma-ggh-wz}, assuming $\cos(\beta-\alpha)=-0.1$. For $WW \ (ZZ)$ decay channel, the LHC has excluded the CP-even Higgs with masses up to 360 (390) GeV and $\tan\beta$ below 1 (3).
With realistic branching fractions at $\tan\beta=10 \ (1)$, the 27 TeV LHC may discover the CP-even Higgs as heavy as 1.1 TeV ($1.5-2$ TeV) through $gg\to H^0\to W^+W^-, ZZ$ channels as shown in Fig.~\ref{sigma-ggh-wz} (b) and (d). The loss of sensitivity at large $\tan\beta$ is mainly due to the reduction of BR$(H^0\to W^+W^-, ZZ)$.

\section{Single Charged Higgs Production}
\label{sec:singleHpm}

If the charged Higgs boson is heavier than the top quark mass, the conventional production of heavy charged Higgs is through $gg\to t b H^\pm$. However, at high energy colliders, an ordinary $p_T$ cut (several tens of GeV) on the $b$-jet in final states is not enough as ${\rm log}(\sqrt{\hat{s}}/p_T)$ is still very large. Thus, this exclusive contribution is only meaningful when detecting final state $b$-jet with sufficiently large $p_T$ cut as regulator. A more dominant mode would be taking $b$ as a parton and considering ``inclusive'' production. Thus, the leading production mechanism would be the associated production of $H^\pm$ with a top quark~\cite{Gunion:1986pe,Akeroyd:2016ymd}
\beq
g b \to t H^\pm .
\eeq
Its total cross section is more accurately estimated~\cite{Belyaev:2002eq,Berger:2003sm,Flechl:2014wfa}.

\begin{figure}[t]
\begin{center}
\minigraph{7cm}{-0.05in}{(a)}{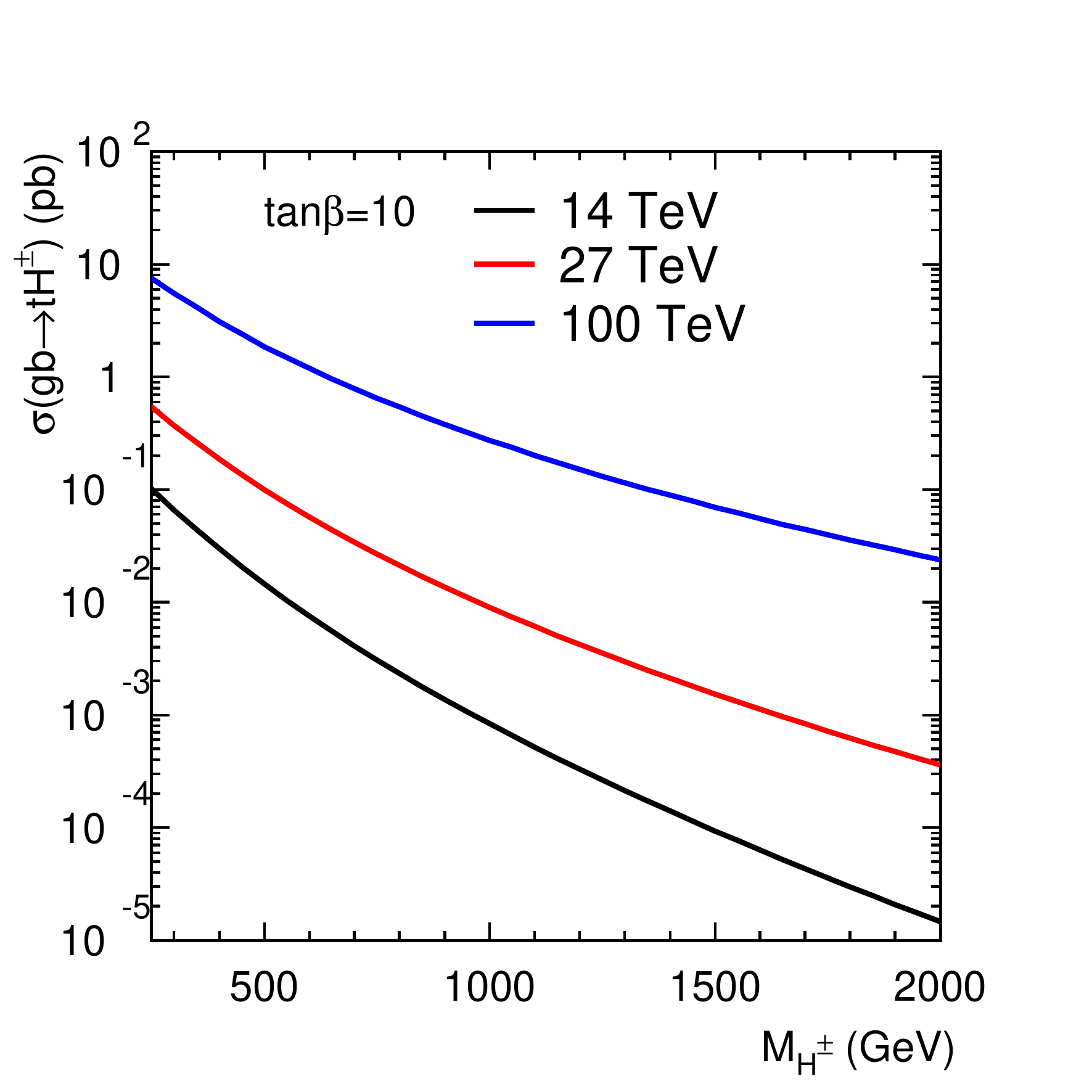}
\minigraph{7cm}{-0.05in}{(b)}{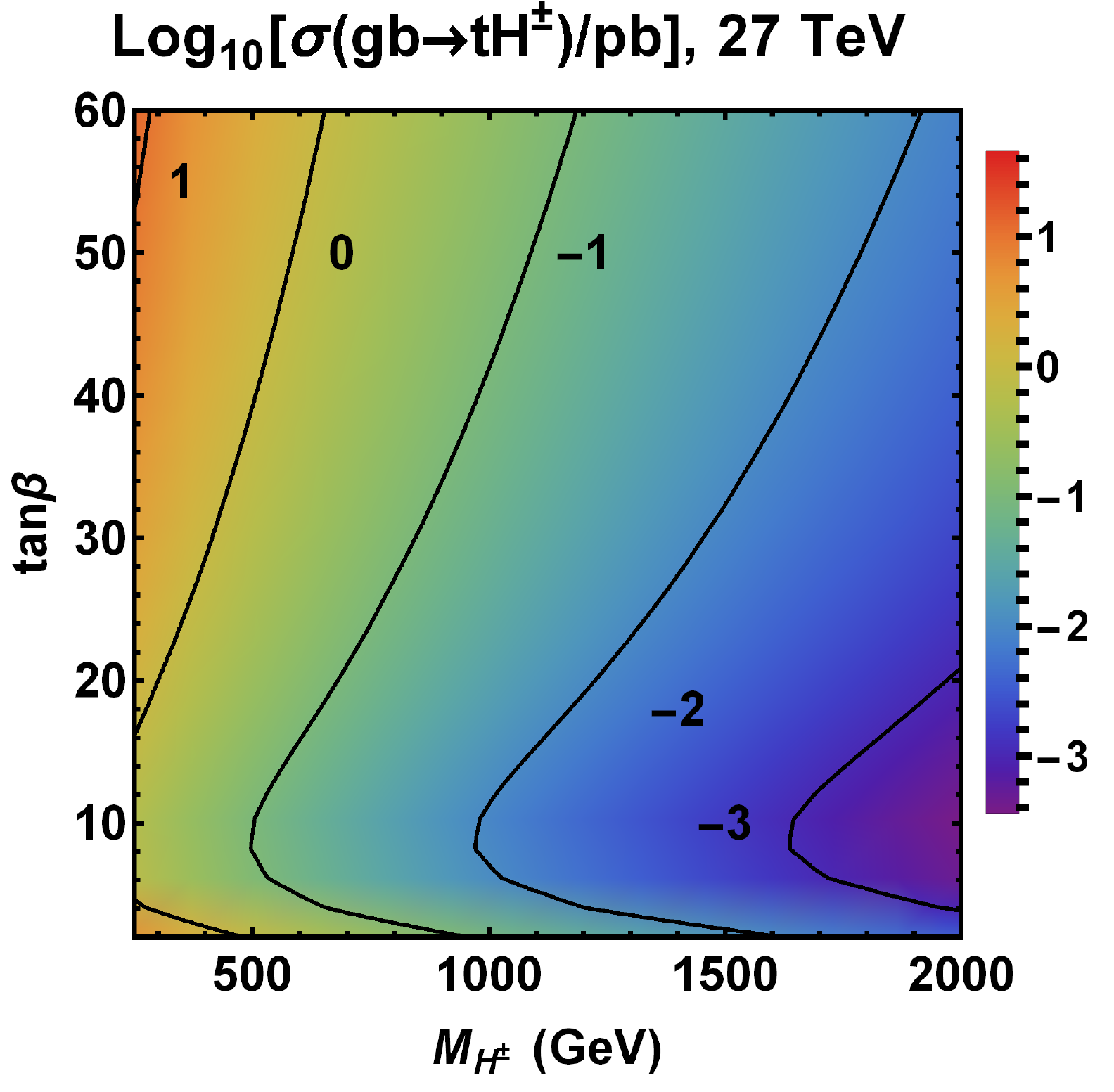}
\end{center}
\caption{Left: Total production cross section versus the Higgs boson mass for $g b\to t H^\pm$ with $\tan\beta=10$ at $pp$ collider with 14 TeV, 27 TeV and 100 TeV. Right: The cross section indicated by contour lines in the plane of $\tan\beta$ versus the Higgs boson mass for $g b\to t H^\pm$ at the 27 TeV LHC. }
\label{fig:xsecHpmt}
\end{figure}

The production cross sections versus charged Higgs mass are shown in Fig.~\ref{fig:xsecHpmt} at the 14 TeV LHC, 27 TeV LHC, as well as the 100 TeV collider. They are the leading order results with a running bottom quark Yukawa coupling at the scale of the pole mass $m_b=4.6$ GeV.
The total production cross section at 27 TeV LHC ranges from 0.5 pb at $M_{H^\pm}=250$ GeV to $4 \times 10^{-4}$ pb at $M_{H^\pm}=2$ TeV for $\tan\beta=10$.
We quantify the signal observability according to the leading decay channels.

\subsection{$ H^\pm \to \tau^\pm \nu $}
\label{sec:Htaunu}
We first consider the clean channel of the charged Higgses' leptonic decay, i.e. $H^\pm \to \tau^\pm \nu$ with $\tau^\pm\to \pi^\pm \nu$,
and the hadronic decay of the $W$ boson from the top quark.
This channel with $\tau$ lepton has been studied before and it was argued to be a good production mode for the LHC energy upgrade to search for~\cite{Basso:2015dka,Aboubrahim:2018tpf}.
We adopt the basic acceptance cuts
\begin{eqnarray}
&&p_T(\ell)\geq 30 \ {\rm GeV}, \ \ \ p_T(b, \pi)\geq 25 \ {\rm GeV}; \ \ \ |\eta(b, \pi, \ell)|<2.5; \ \ \ \Delta R\geq 0.4.
\label{basic1}
\end{eqnarray}
The leading SM backgrounds are given by $gb\to W^\pm t$ with $W^\pm\to \tau^\pm \nu_\tau$ and QCD $t\bar{t}$ production with one $b$-jet being vetoed if
$p_T(b)>30 \ {\rm GeV}, |\eta(b)|<4.9$.

\begin{figure}[h!]
\begin{center}
\minigraph{7cm}{-0.05in}{(a)}{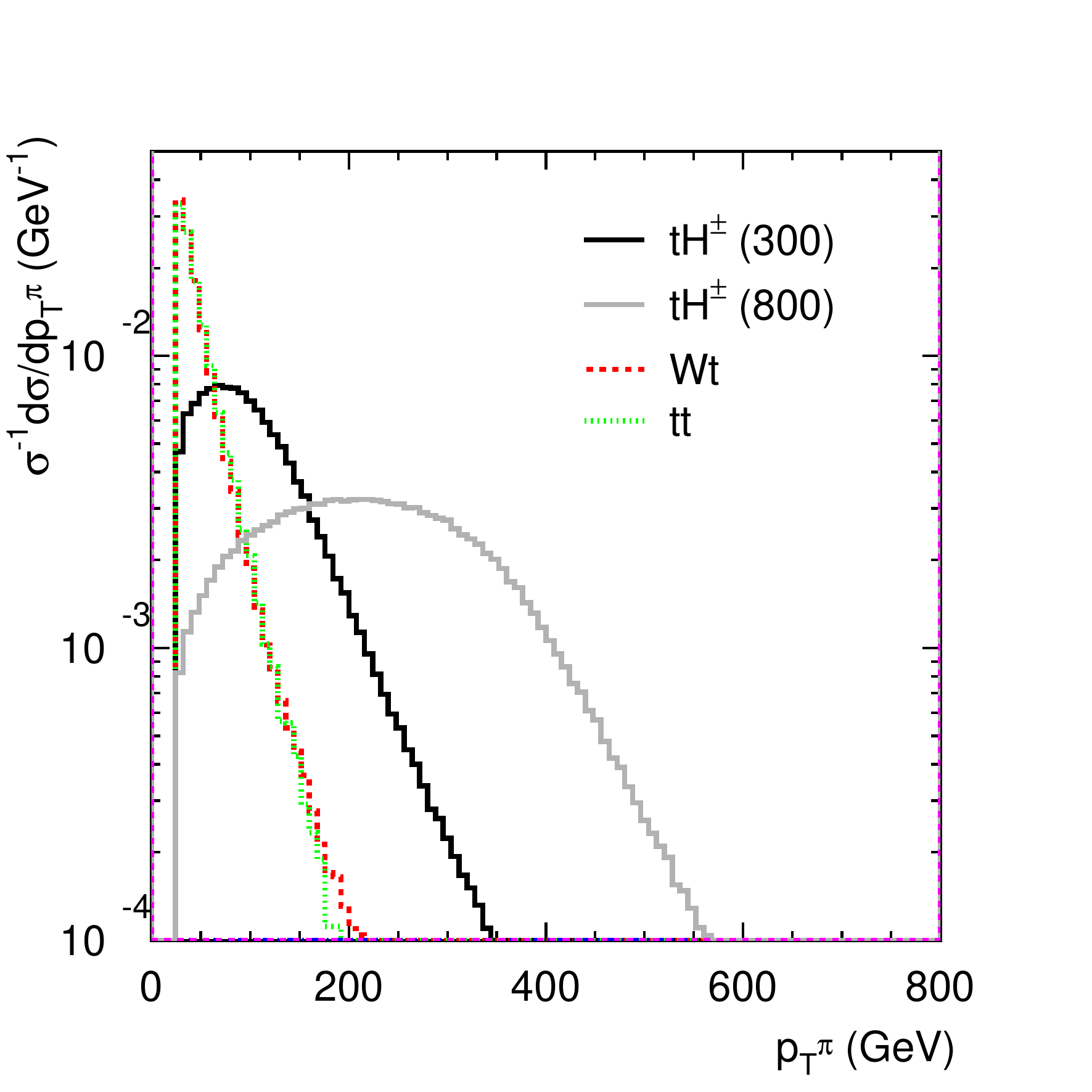}
\minigraph{7cm}{-0.05in}{(b)}{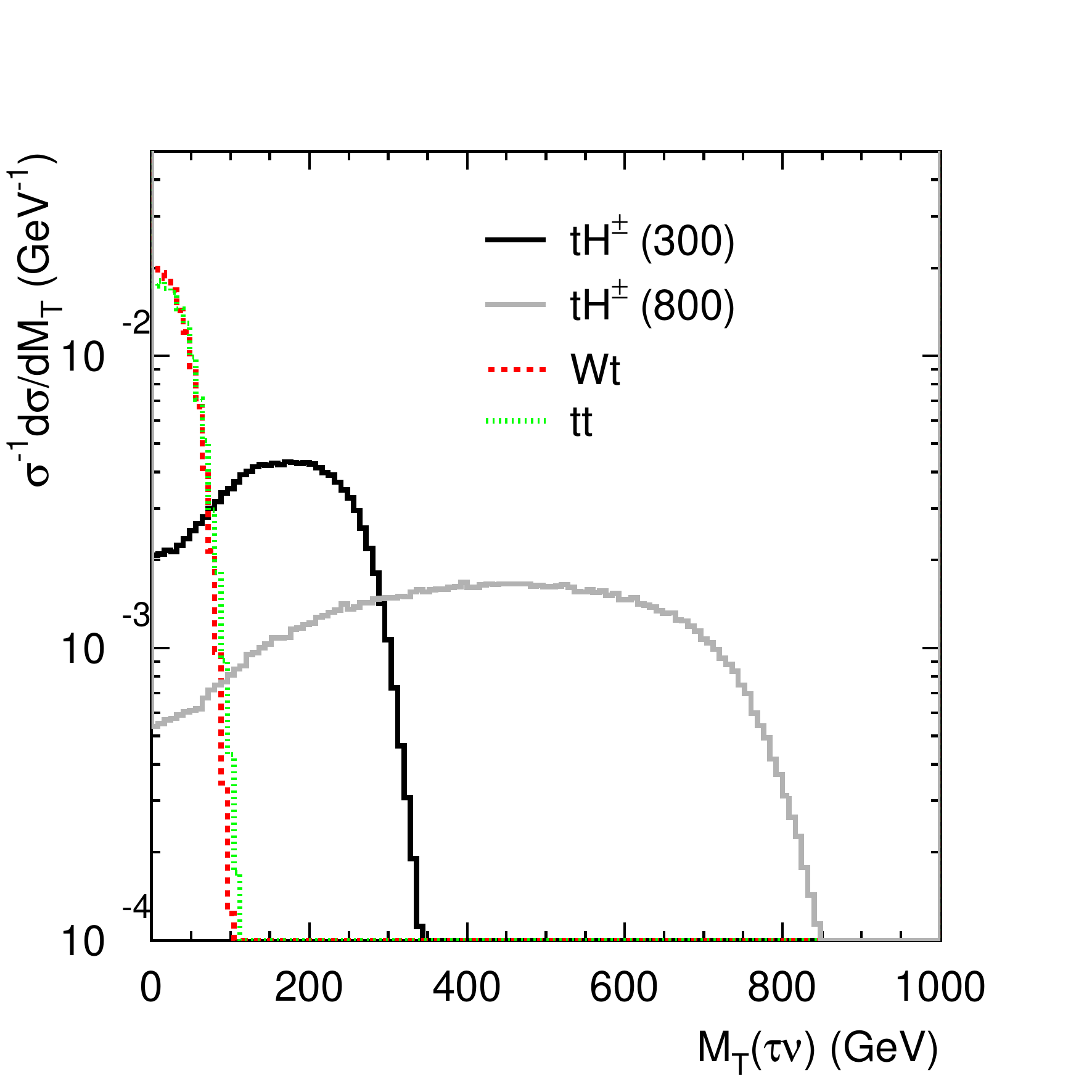}
\end{center}
\caption{The differential cross section distributions of $p_T(\pi)$ (a) and $M_T(\tau\nu)$ (b) for the signal $gb\to t H^\pm\to \tau^\pm \nu b W^\mp\to \tau^\pm \nu bjj$ and backgrounds at the 27 TeV LHC. }
\label{hpmt-taunu-jdis}
\end{figure}

Note that, as the charged Higgs $H^{-}$ only coupled with right-handed charged lepton, the right-handed $\tau^-_R$ decays to a left-handed $\nu_\tau$ and $\pi^-$. This causes the $\pi^-$ to preferentially move along the $\tau^-$ momentum direction. In contrast, the
$\tau^-$ coming from $W^{-}$ decay is left-handed, which has the opposite effect on the $\pi^-$. The similar feature holds for the $\tau^+$ from $H^{+}$ and $W^+$ decays. This is a well-known result of spin
correlation in the $\tau$ decay~\cite{Bullock:1991fd,Bullock:1992yt}. Thus, the transverse momentum of $\pi^\pm$ from charged Higgs decay to tau lepton yields a harder spectrum than that from $W$ decay in SM backgrounds~\cite{Cao:2003tr,Christensen:2012si,Li:2015lra}, as seen in Fig.~\ref{hpmt-taunu-jdis} (a).
We thus tighten the missing energy and the $p_T$ of pion
\begin{eqnarray}
\cancel{E}_T>100 \ {\rm GeV}, \ \ \ p_T(\pi)>65 \ {\rm GeV}.
\end{eqnarray}
Furthermore, Fig.~\ref{hpmt-taunu-jdis} (b) indicates that the transverse mass of the pion and missing neutrinos from charged Higgs
\begin{eqnarray}
M_{T}(\tau\nu)=\sqrt{(p_T(\pi)+\cancel{E}_T)^2-(\vec{p}_T(\pi)+\vec{\cancel{p}}_T)^2}
\end{eqnarray}
should be greater than 100 GeV in order to reduce backgrounds. One can see that these cuts help reduce the backgrounds significantly from the cut efficiencies shown in Table~\ref{cuteff-hpmt-taunu-j}.

\begin{table}[tb]
\begin{center}
\begin{tabular}{|c|c|c|c|c|}
\hline
cut efficiencies  & basic cuts & $\cancel{E}_T$ & $p_T^\pi$ & $M_T$
\\ \hline
$t H^\pm (300)$ & 0.36 & 0.22 & 0.16  & 0.14  \\
$t H^\pm (800)$ & 0.40 & 0.36 & 0.34  & 0.33   \\
\hline \hline
$Wt$ & 0.1 & 0.034 & 0.0087  & negligible   \\
\hline
$t\bar{t}$ & 0.026 & 0.012 & 0.0026 & $5\times 10^{-6}$   \\
\hline
\end{tabular}
\end{center}
\caption{The cut efficiencies for $gb\to t H^\pm\to \tau^\pm \nu b W^\mp\to \tau^\pm \nu bjj$ and the SM backgrounds after consecutive cuts at the 27 TeV LHC. We take $M_{H^{\pm}}=300$ or 800 GeV.}
\label{cuteff-hpmt-taunu-j}
\end{table}

\begin{figure}[h!]
\begin{center}
\minigraph{7cm}{-0.05in}{(a)}{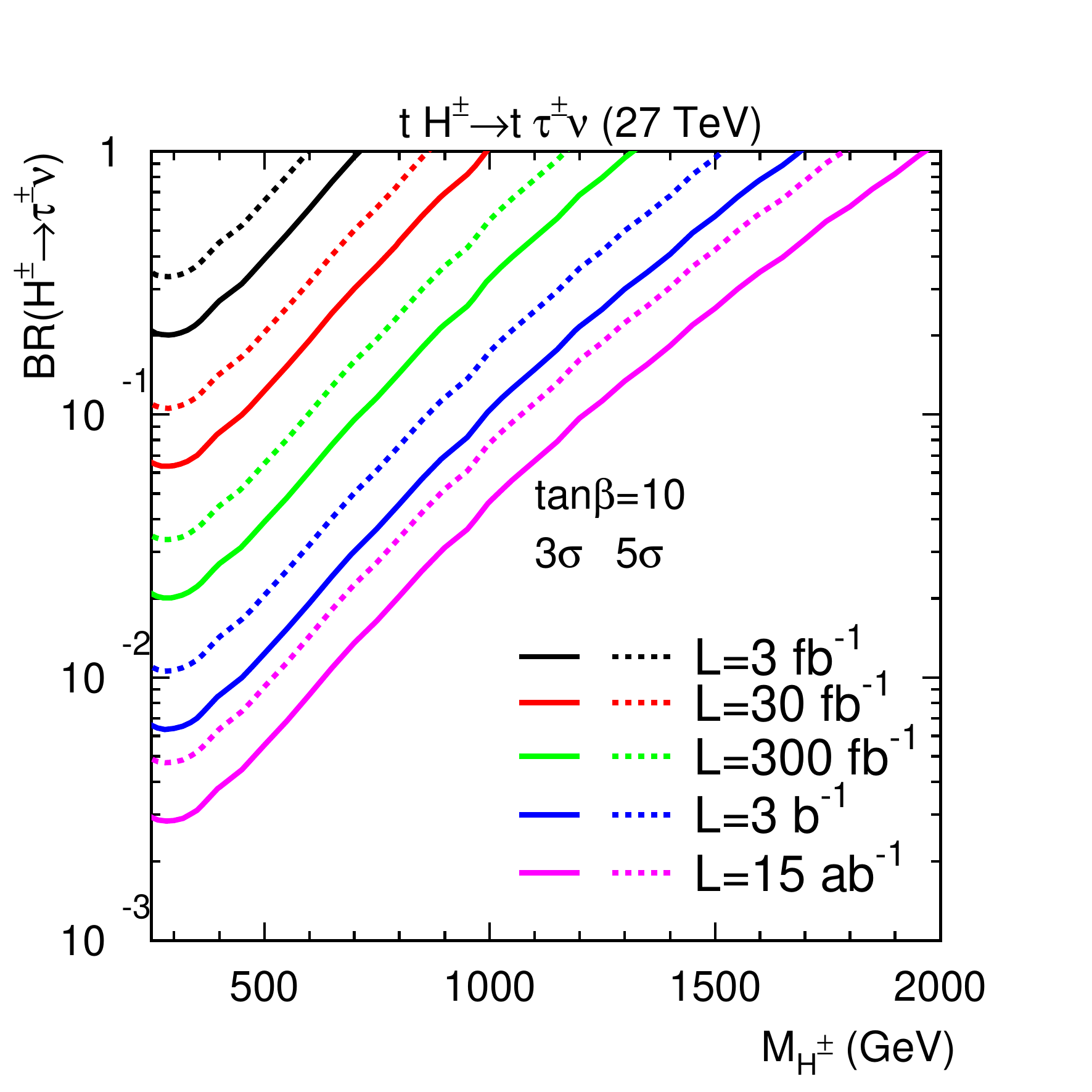}
\minigraph{6.2cm}{-0.05in}{(b)}{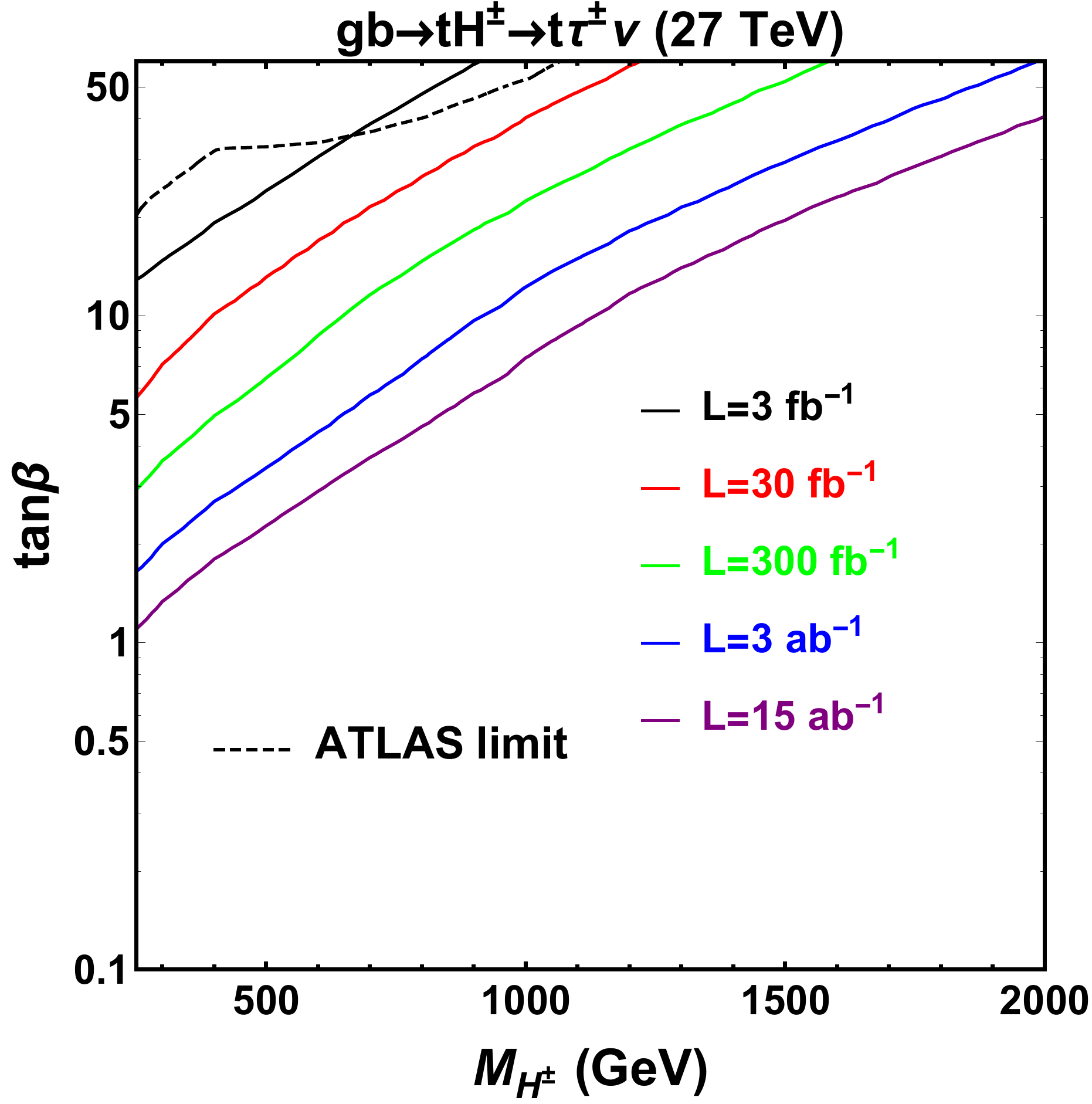}
\end{center}
\caption{
Left: Reach of BR$(H^\pm\to \tau^\pm\nu)$ as a function of $M_{H^\pm}$ for $gb\to t H^\pm\to \tau^\pm \nu bjj$ channel at the 27 TeV LHC. We assume $\tan\beta=10$.
Right: Discovery contour in $\tan\beta$ versus $M_{H^\pm}$ plane for $gb\to t H^\pm\to \tau^\pm \nu bjj$ with realistic BR$(H^\pm\to \tau^\pm\nu)$.
As a comparison, the 13 TeV LHC exclusion limit on $\tan\beta$ as a function of $M_{H^\pm}$ is also presented~\cite{Aaboud:2018gjj}.
}
\label{sigma-hpmt-taunu-j}
\end{figure}

If the exotic decay modes (one neutral Higgs with $W$ boson) are absent, the charged Higgs decay is actually dominated by $tb$ mode once it is kinematically open. The $H^\pm\to \tau^\pm\nu$ decay is the secondary significant mode in the decays to the SM particles and becomes more important as $\tan\beta$ increases. Figure~\ref{sigma-hpmt-taunu-j} (a) displays the reachable limit of BR$(H^\pm\to \tau^\pm\nu)$ at the 27 TeV LHC. The HE-LHC with 15 ab$^{-1}$ luminosity extends the reach of BR$(H^\pm\to \tau^\pm\nu)$ to $10^{-3}$ level for $\tan\beta=10$.

The 13 TeV LHC performed the search for charged Higgs bosons through the production of a heavy charged Higgs boson in association with $t$ and $b$ quarks~\cite{Aaboud:2018gjj,Sirunyan:2019hkq}. The results are interpreted in the framework of the hMSSM scenario which is a Type-II 2HDM~\cite{Djouadi:2013uqa}. As a comparison, the 95\% CL exclusion limit on $\tan\beta$ as a function of $M_{H^\pm}$ is also presented in Fig.~\ref{sigma-hpmt-taunu-j} (b). The charged Higgs boson mass is excluded up to 1.1 TeV for $\tan\beta=60$, with the integrated luminosity of 36 fb$^{-1}$~\cite{Aaboud:2018gjj}.
With realistic BR$(H^\pm\to \tau^\pm\nu)$, the discovery region in $\tan\beta$ versus $M_{H^\pm}$ plane is shown in Fig.~\ref{sigma-hpmt-taunu-j} (b) for $gb\to t H^\pm\to \tau^\pm \nu bjj$ channel at 27 TeV LHC. The region below $\tan\beta\sim 1$ can not be covered by $5\sigma$ discovery due to the suppression of the decay branching fraction. The 27 TeV pp collider with 3 ab$^{-1}$ luminosity can discover the charged Higgs mass up to 1 TeV (2 TeV) for $\tan\beta=10 \ (60)$.

\subsection{$ H^\pm \to tb $}
\label{sec:Htb}

Next we consider the signal induced by decay $H^\pm \to tb$ followed by the two top quarks' semi-leptonic decays, i.e. $gb\to t H^\pm\to b t\bar{t}\to bbbjj\ell^\pm \nu$. The irreducible SM background is thus $gb\to bt\bar{t}$. The basic cuts are the same as those in Eq.~(\ref{basic1}) for jets and lepton.
Any $b$-jets in the events are assumed to be tagged with an efficiency of 70\%.

\begin{figure}[h!]
\begin{center}
\includegraphics[scale=1,width=7cm]{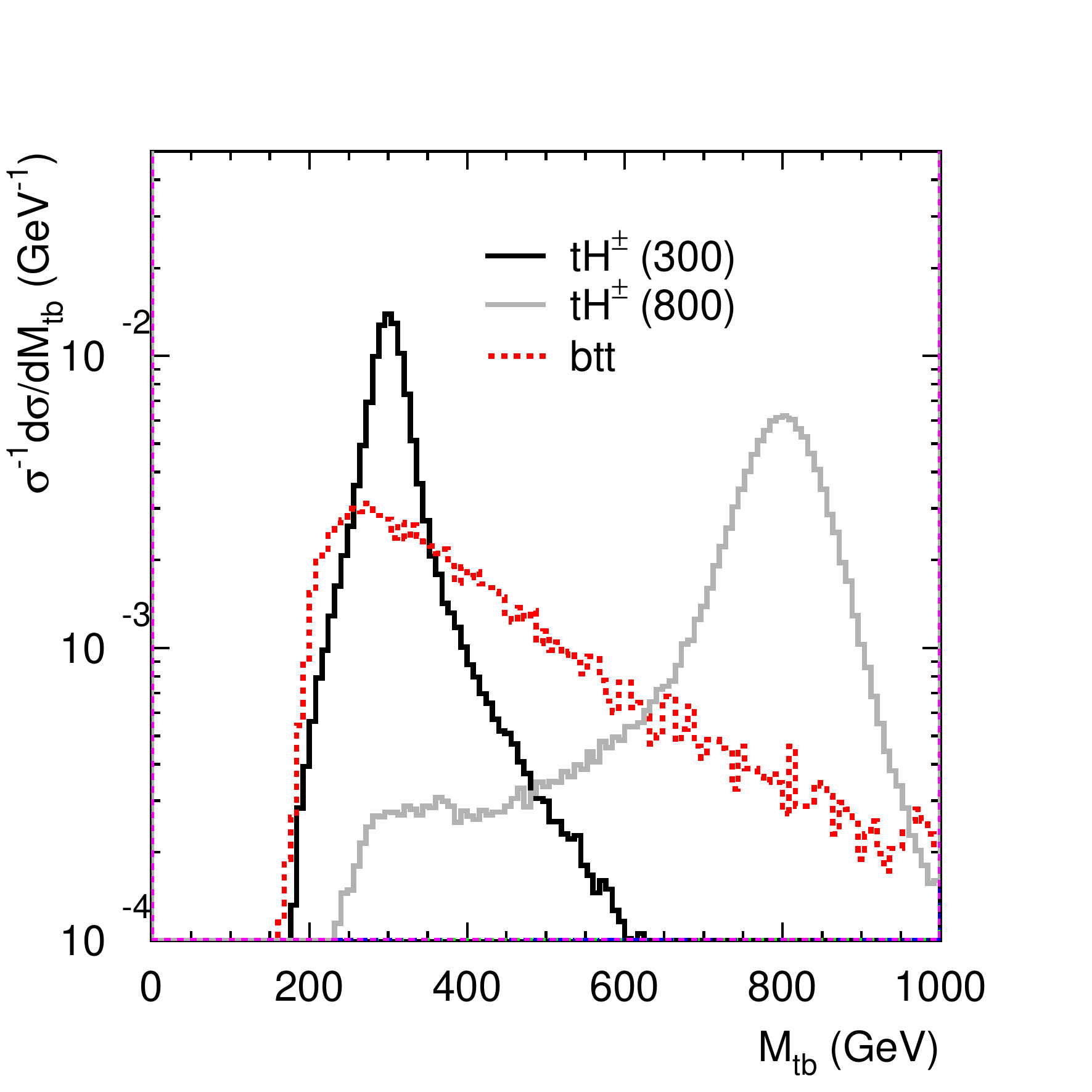}
\end{center}
\caption{The differential cross section distribution of the invariant mass $M_{tb}$ for the signal $gb\to t H^\pm\to b t\bar{t}\to bbbjj\ell^\pm \nu$ and backgrounds at the 27 TeV LHC. }
\label{hpmt-tb-jdis}
\end{figure}

As the missing neutrino is only from $W$'s leptonic decay, using $W$'s mass and the missing transverse momentum $\cancel{\vec{p}}_T$, one can arrive at a solution of the longitudinal momentum of the neutrino and this $W$ boson can thus be reconstructed~\cite{Cao:2003tr}. The other $W$ can be directly reconstructed by the invariant mass of the two light jets. The three $b$-jets are then assigned with the two $W$ bosons to fully reconstruct two top quarks and the charged Higgs. The invariant mass of $tb$ for the charged Higgs is displayed in Fig.~\ref{hpmt-tb-jdis}. We apply the kinematic cuts on the missing energy, the $p_T$ of $b$-jet and the invariant mass of charged Higgs as follows
\begin{eqnarray}
\cancel{E}_T>40 \ {\rm GeV}, \ \ \ p_T^{\rm max}(b)>M_{H^\pm}/3, \ \ \ |M_{tb}-M_{H^\pm}| < M_{H^\pm}/10.
\end{eqnarray}
The resultant cut efficiencies are listed in Table~\ref{cuteff-hpmt-tb-j}. Due the the complexity of the objects in final states, it turns out that these cuts are not as efficient as those for $H^\pm\to \tau^\pm \nu$ signal.

\begin{table}[tb]
\begin{center}
\begin{tabular}{|c|c|c|c|c|}
\hline
cut efficiencies  & basic cuts & $\cancel{E}_T$ & $p_T^b$ & $M_{tb}$
\\ \hline
$t H^\pm (300)$ & 0.18 & 0.11 & 0.075 & 0.045 \\
$t H^\pm (800)$ & 0.22 & 0.18 & 0.14 & 0.1 \\
\hline \hline
$btt (300)$ & 0.015 & 0.0096 & 0.0064 & 0.00075 \\
$btt (800)$ & 0.015 & 0.0096 & 0.0011 & 0.00019 \\
\hline
\end{tabular}
\end{center}
\caption{The cut efficiencies for $gb\to t H^\pm\to b t\bar{t}\to bbbjj\ell^\pm \nu$ and the SM backgrounds after consecutive cuts at the 27 TeV LHC. We take $M_{H^{\pm}}=300$ or 800 GeV.}
\label{cuteff-hpmt-tb-j}
\end{table}

\begin{figure}[h!]
\begin{center}
\minigraph{7cm}{-0.05in}{(a)}{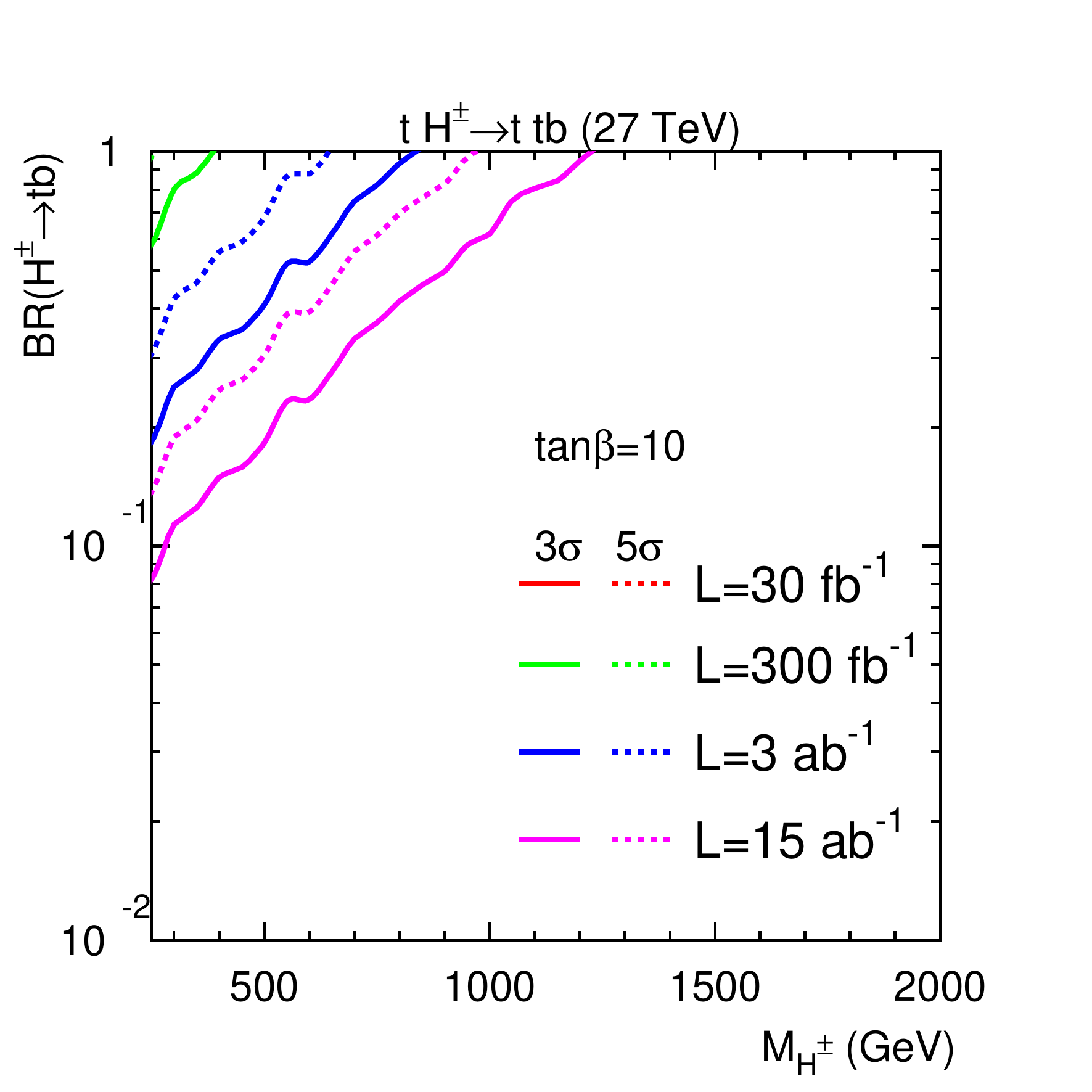}
\minigraph{6cm}{-0.05in}{(b)}{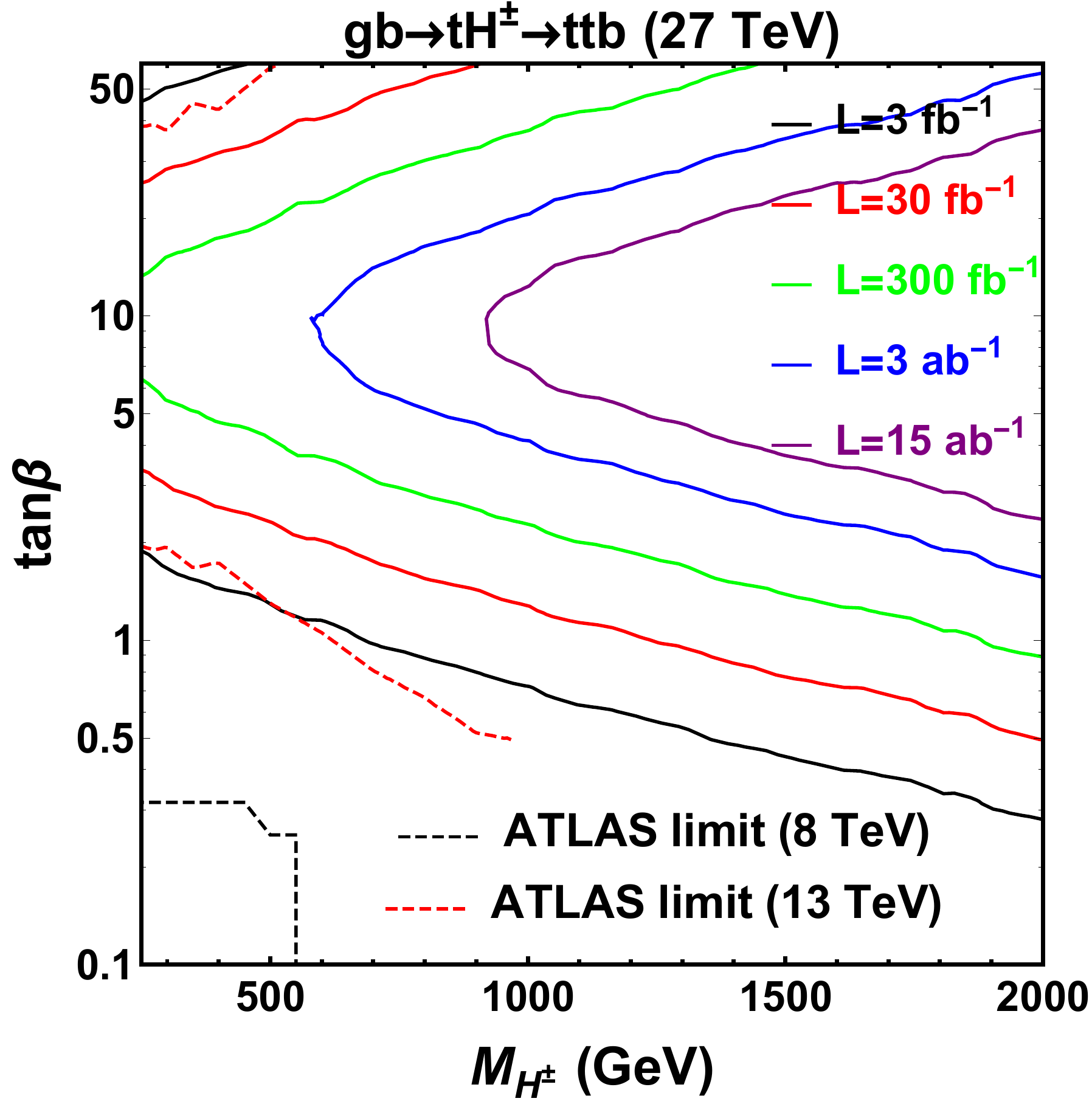}
\end{center}
\caption{
Left: Reach of BR$(H^\pm\to tb)$ as a function of $M_{H^\pm}$ for $gb\to t H^\pm\to b t\bar{t}\to bbbjj\ell^\pm \nu$ channel at the 27 TeV LHC. We assume $\tan\beta=10$.
Right: Discovery contour in $\tan\beta$ versus $M_{H^\pm}$ plane for $gb\to t H^\pm\to b t\bar{t}\to bbbjj\ell^\pm \nu$ with realistic BR$(H^\pm\to tb)$.
The observed exclusion limits at 8 TeV~\cite{Aad:2015typ} and 13 TeV~\cite{Aaboud:2018cwk} LHC are indicated by dashed curves.
}
\label{sigma-hpmt-tb-j}
\end{figure}

The left panel of Fig.~\ref{sigma-hpmt-tb-j} shows the reachable limit of BR$(H^\pm\to tb)$ as a function of $M_{H^\pm}$ with $\tan\beta=10$. With 15 ab$^{-1}$ luminosity, the charged Higgs mass can be probed as heavy as $M_{H^\pm}\simeq 950$ GeV for 5$\sigma$ discovery if BR$(H^\pm\to tb)=1$. As $H^\pm\to tb$ is the leading decay mode for both small and large $\tan\beta$ in the alignment limit, this discovery potential is also true for realistic values of BR$(H^\pm\to tb)$ calculated by 2HDMC as shown in Fig.~\ref{sigma-hpmt-tb-j} (b). In Fig.~\ref{sigma-hpmt-tb-j} (b), the regions to the left of the curves are covered by 5$\sigma$ discovery at 27 TeV LHC. One can see that final states with $t H^\pm\to bt\bar{t}$ prove to be a very sensitive channel for regions with both small and large $\tan\beta$. Although the region with moderate $\tan\beta$ loses sensitivity due to the suppression of the decay branching fraction, the charged Higgs with mass up to 900 GeV can be probed for $\tan\beta\simeq 10$ with 15 ab$^{-1}$ luminosity.

Early phenomenological studies have performed the analysis of this signature and concluded that the LHC discovery potential might be optimistic for the charged Higgs mass lower than 600 GeV~\cite{Barger:1993th,Gunion:1993sv,Miller:1999bm,Moretti:1999bw,Pedersen:2016kyw}.
The LHC explored heavy charged Higgs boson decaying into $t\bar{b}(\bar{t}b)$ through $gb\to t H^\pm$ at $\sqrt{s}=8$ TeV~\cite{Aad:2015typ} and $gg\to tb H^\pm$ at $\sqrt{s}=13$ TeV~\cite{Aaboud:2018cwk,Sirunyan:2019hkq}. We convert the observed limit on the production cross section $\sigma(gb\to t H^\pm)$ times branching fraction
for $H^\pm\to tb$ to the constraint on $\tan\beta$ versus $M_{H^\pm}$ in Type-II 2HDM, as shown by black dashed curve in Fig.~\ref{sigma-hpmt-tb-j} (b). The 8 TeV LHC excluded charged Higgs mass up to 550 GeV and $\tan\beta$ below 0.3. In the scenario of hMSSM as shown by red curves, at 13 TeV LHC with 36 fb$^{-1}$ luminosity, the observed exclusion for $M_{H^\pm}$ is in the range $200-965$ GeV for $0.5<\tan\beta<1.95$ and high values of $\tan\beta$ between 36 and 60 are excluded in the $M_{H^\pm}$ range $220-540$ GeV.

\section{Pair Production of Higgs Bosons}
\label{sec:pairHiggs}

\begin{figure}[t]
\begin{center}
\minigraph{7cm}{-0.05in}{(a)}{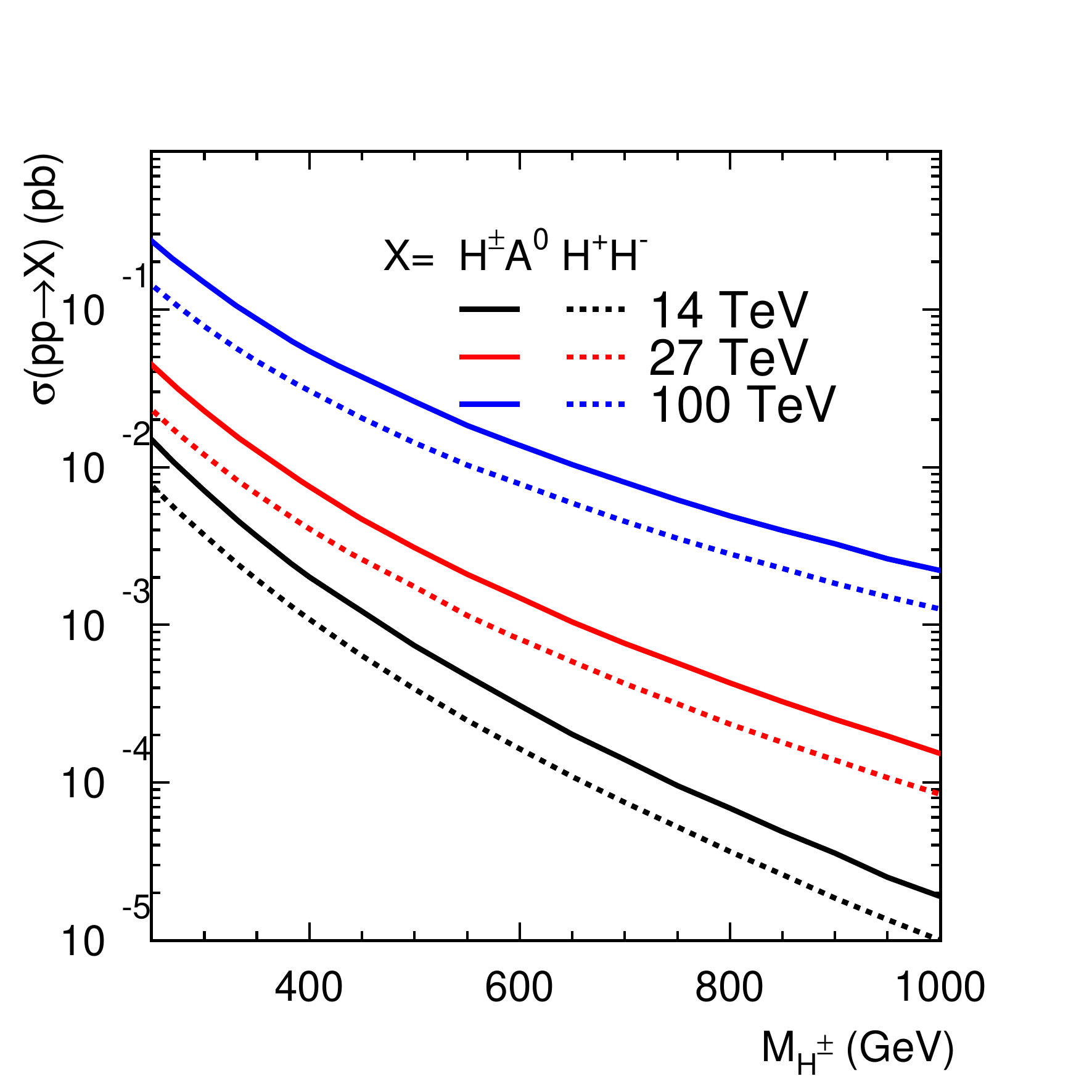}
\minigraph{7cm}{-0.05in}{(b)}{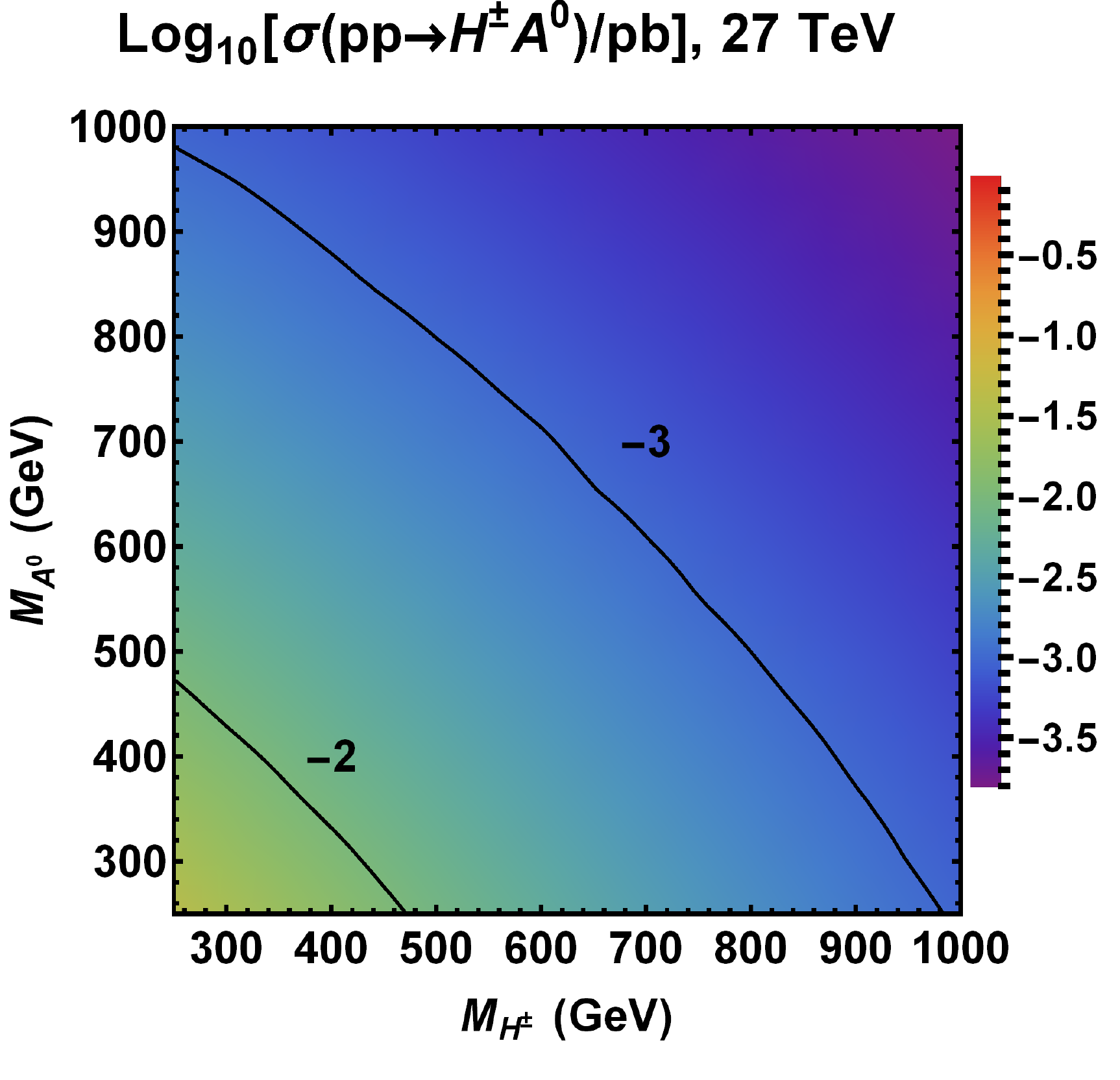}
\end{center}
\caption{Left: Total production cross section versus the Higgs boson mass for $q\bar q \to H^\pm A^0,\ H^+ H^-$
with $M_{A^0}=M_{H^\pm}$ at $pp$ collider with 14 TeV, 27 TeV and 100 TeV. Right: The cross section indicated by contour lines in the plane of $M_{A^0}$ versus $M_{H^\pm}$ for $q\bar q \to H^\pm A^0$ at the 27 TeV LHC. }
\label{fig:xsecHiggsPair}
\end{figure}

Besides the above leading production channels of single Higgs boson, the electroweak production of Higgs boson pairs are potentially important. Their total production cross sections are independent of any model parameters except for Higgs masses as they are via pure electroweak gauge interactions.
The pair productions of Higgs bosons through pure gauge interactions are~\cite{Kanemura:2001hz,Cao:2003tr,Christensen:2012ei,Christensen:2012si,Dawson:2012gs}
\begin{eqnarray}
q\bar{q}'\to W^{\pm\ast}\to H^\pm A^0, \ \ \ q\bar{q}\to Z^\ast/\gamma^\ast\to H^+H^- .
\end{eqnarray}
The relevant Higgs couplings to gauge bosons scale as
\begin{eqnarray}
WH^\pm A^0\propto g/2, \ ZH^+H^-\propto -g \cos 2\theta_W/(2c_W), \ \gamma H^+H^-\propto -ie ,
\end{eqnarray}
where $g$ is the weak coupling and $\theta_W$ is the weak-mixing angle with $c_W = \cos\theta_W$.
Figure~\ref{fig:xsecHiggsPair} shows their total cross sections at 14 TeV LHC, 27 TeV LHC and 100 TeV $pp$ collider.
The total cross section of $H^\pm A^0$ production at 27 TeV LHC ranges from $2.3\times 10^{-2}$ pb at $M_{A^0}=M_{H^\pm}=250$ GeV to $1.5 \times 10^{-4}$ pb with 1 TeV Higgs mass. It is larger than that of $H^+H^-$ production by about twice.
We explore their observability based on the leading decay modes.

\subsection{$H^\pm A^0\to \tau^\pm\nu b\bar{b}$}
\label{HpmAtau}

The first signal channel we consider is the associated production of the CP-odd Higgs $A^0$ and the charged Higgs $H^\pm$, followed by $A^0$ and $H^\pm$ decay to  $b\bar{b}$ and $\tau^\pm \nu_\tau$ respectively, i.e. $pp\to H^{\pm}A^{0}\to \tau^\pm \nu_\tau b\bar{b}$.
We again adopt the $\tau$'s leading 2-body decay channel, i.e. $\tau^\pm\to \pi^\pm \nu_\tau$, with the branching fraction being ${\rm BR}(\tau^\pm\to \pi^\pm \nu_\tau)=0.11$. 
The $b$-jets and the charged pions $\pi^\pm$ in final states satisfy the following basic cuts
\begin{eqnarray}
p_T(b,\pi)\geq 25 \ {\rm GeV}; \ \ \ |\eta(b,\pi)|<2.5; \ \ \ \Delta R_{bb}, \Delta R_{b\pi}\geq 0.4,
\end{eqnarray}
and any $b$-jets in the events are assumed to be tagged with an efficiency of 70\%.
The major SM backgrounds are thus from the following irreducible contributions
\begin{itemize}
\item the gluon splitting process: $q\bar{q}'\to gW^\pm\to b\bar{b}W^\pm\to b\bar{b}\tau^\pm\nu$ ,
\item the single top production: $q\bar{q}'\to W^{\pm\ast}\to b\bar{t}(\bar{b}t)\to b\bar{b}W^\pm\to b\bar{b}\tau^\pm\nu$ ,
\end{itemize}
and the reducible ones
\begin{itemize}
\item the $W^\pm$-gluon fusion process with a forward jet: $gq\to gq'W^{\pm\ast}\to q'b\bar{t}(\bar{b}t)\to q'b\bar{b}W^\pm\to q'b\bar{b}\tau^\pm\nu$ ,
\item the QCD $t\bar{t}$ production: $t\bar{t}\to b\bar{b}W^+W^-\to b\bar{b}\tau^\pm \ell^\mp \nu's \ (\ell=e,\mu)$.
\end{itemize}
The last two processes having additional jet or lepton can be vetoed by requiring the extra objects with
\begin{eqnarray}
p_T(j)>30 \ {\rm GeV}, |\eta(j)|<4.9; \ \ \ p_T(\ell)>7 \ {\rm GeV}, |\eta(\ell)|<3.5.
\label{veto}
\end{eqnarray}

\begin{figure}[h!]
\begin{center}
\minigraph{7cm}{-0.05in}{(a)}{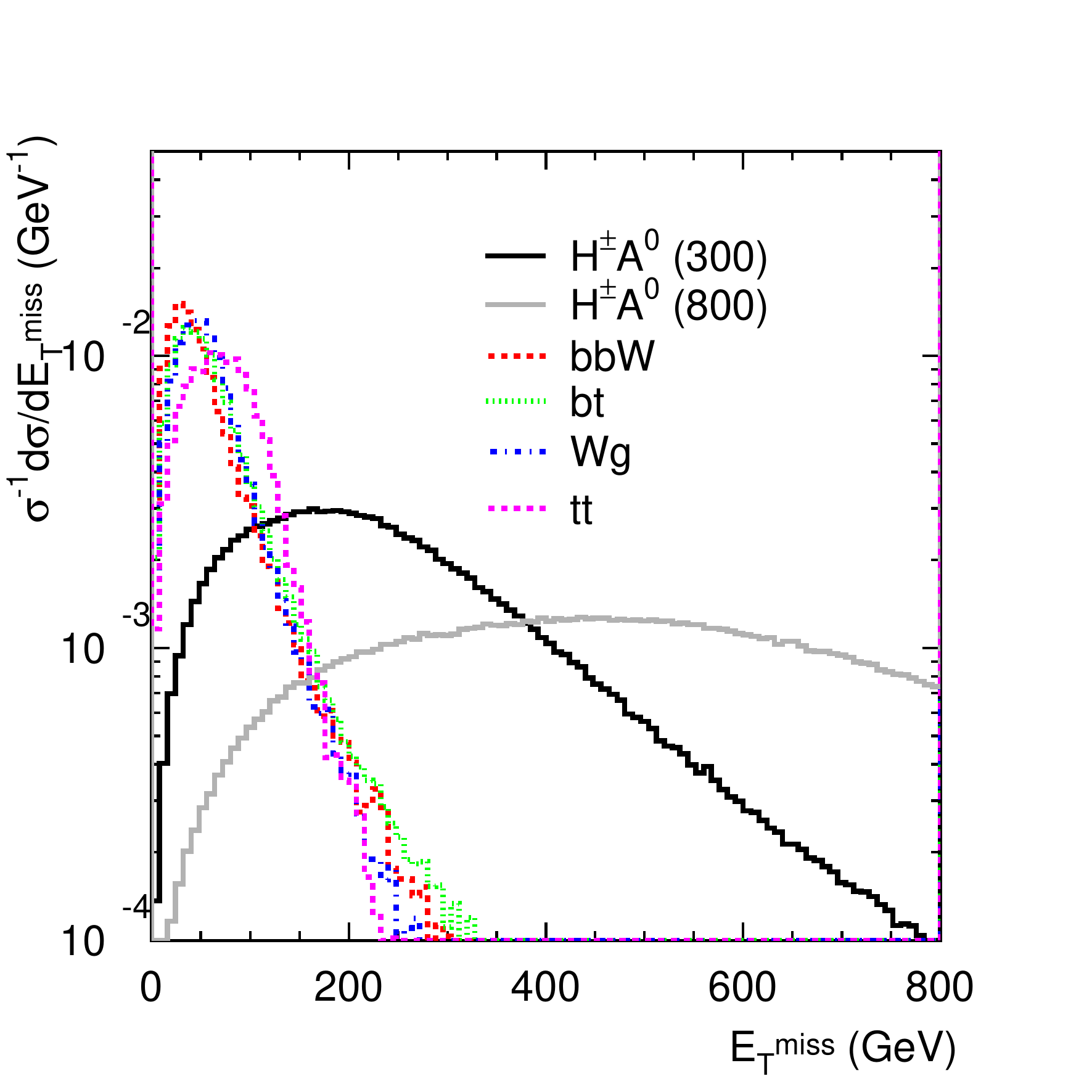}
\minigraph{7cm}{-0.05in}{(b)}{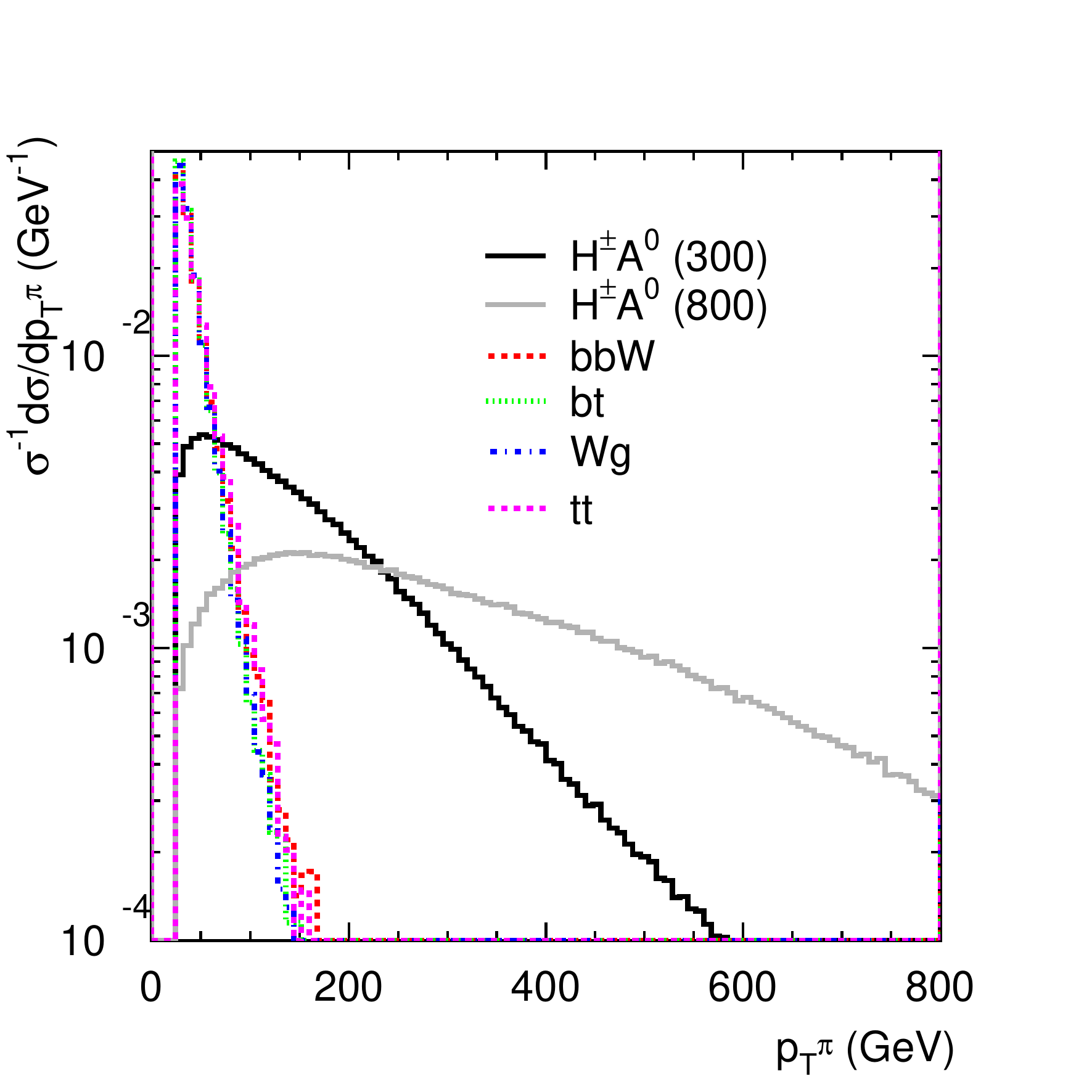}\\
\minigraph{7cm}{-0.05in}{(c)}{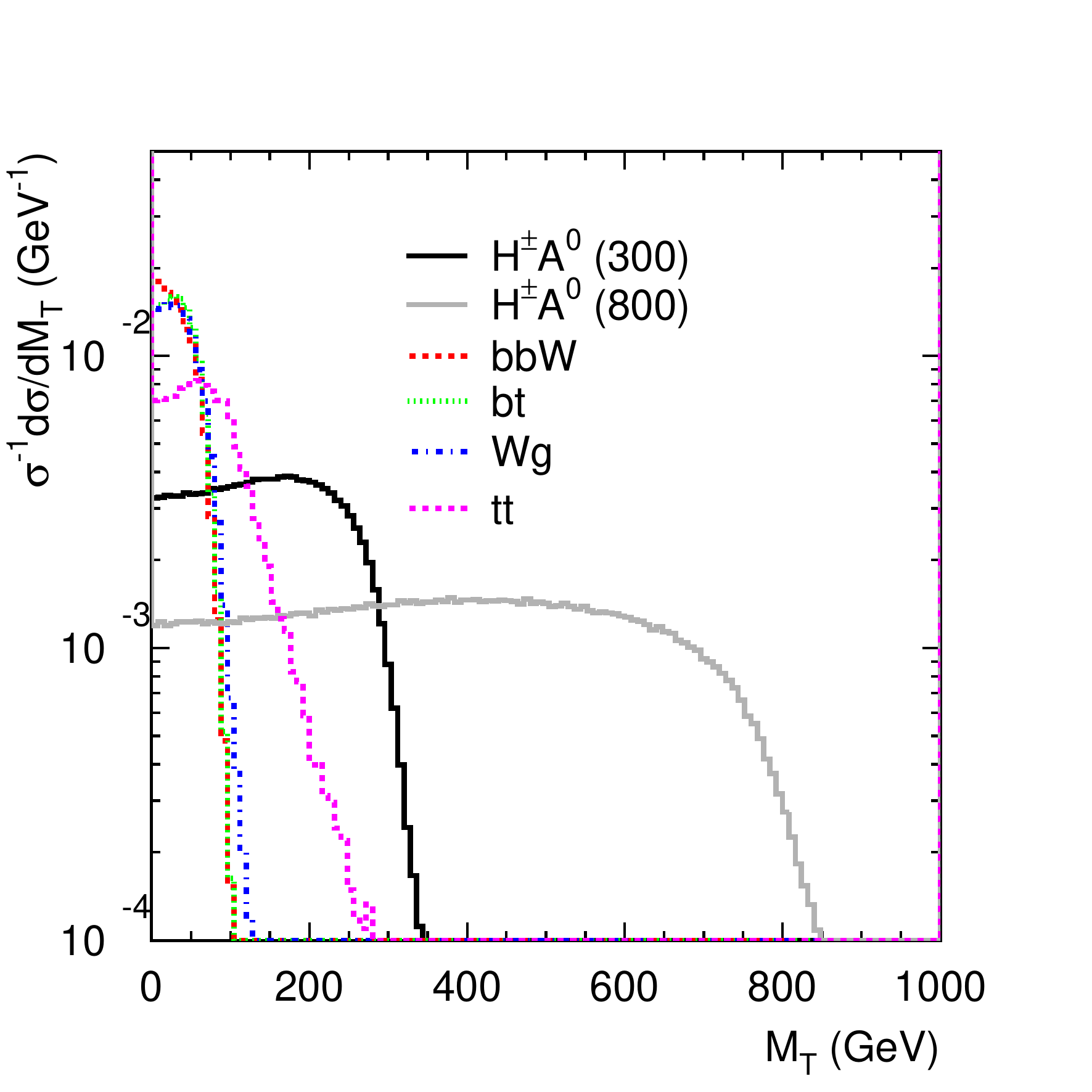}
\minigraph{7cm}{-0.05in}{(d)}{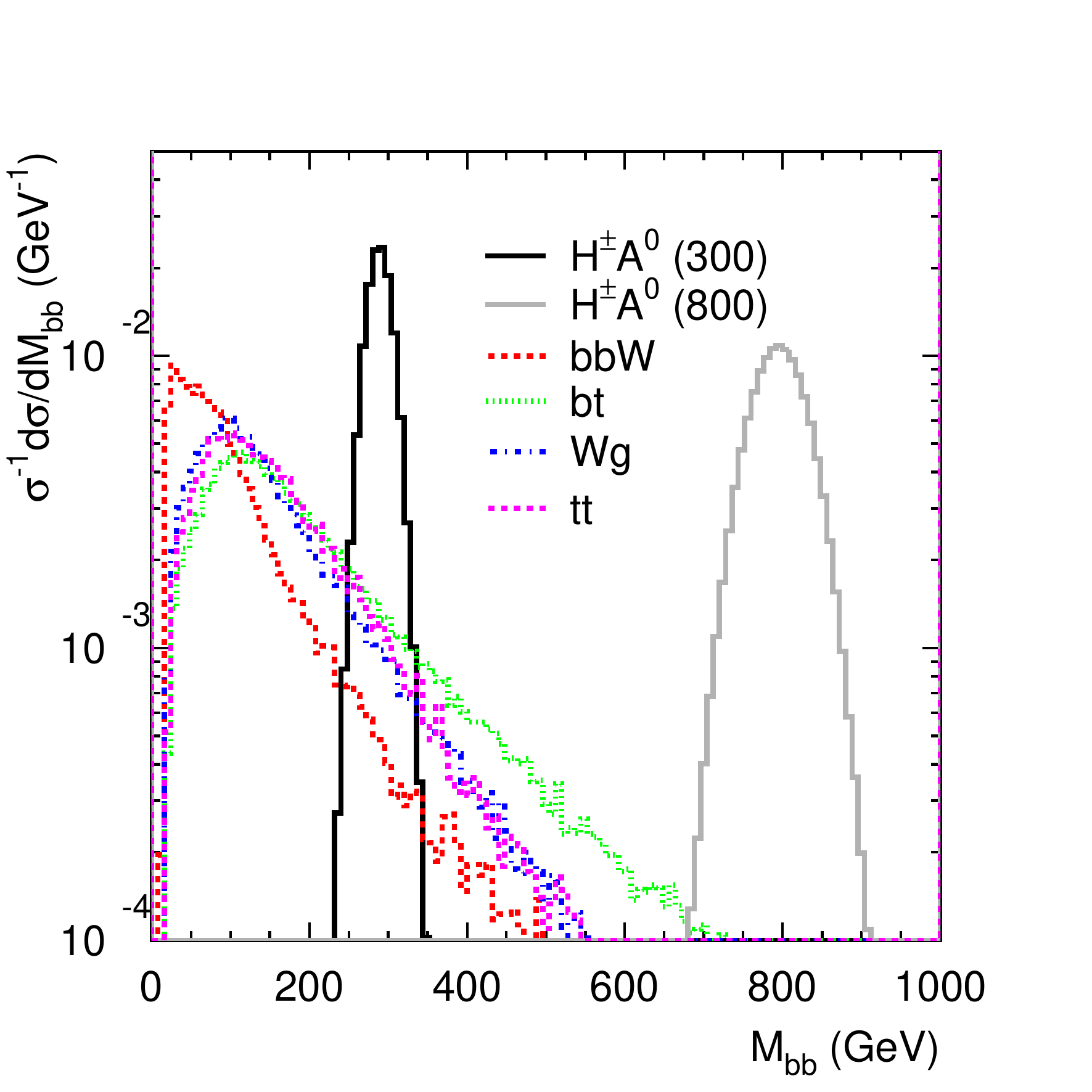}
\end{center}
\caption{The differential cross section distributions of $\cancel{E}_T$ (a), $p_T(\pi)$ (b), $M_T(H^\pm)$ (c) and $M_{bb}$ (d) for the signal $pp\to H^{\pm}A^{0}\to \tau^\pm \nu_\tau b\bar{b}$ and SM backgrounds versus at the 27 TeV LHC. }
\label{h3hpm-dis}
\end{figure}

We display the distributions of signal and backgrounds after the basic cuts at the
27 TeV LHC in Fig.~\ref{h3hpm-dis}, 
for (a) missing transverse energy $\cancel{E}_T$ and (b) transverse pion momentum $p_T(\pi)$.
The signal exhibits a harder $\cancel{E}_T$ spectrum than the SM backgrounds from the Jacobian peak around $p_{T\nu}\sim M_{H^\pm}/2$. The mass peak of the resonance $A^0$ also leads to an enhanced distribution near $p_{Tb}\sim M_{A^0}/2$. Furthermore, as discussed for single $H^\pm$ production with $H^\pm \to \tau^\pm \nu$ in Sec.~\ref{sec:Htaunu}, the signal has a harder $p_T$ distribution of $\pi^\pm$ compared to the SM backgrounds.
The charged Higgs mass $M_{H^\pm}$ and the CP-odd Higgs mass $M_{A^0}$ can be read from
the edge of transverse mass
\begin{eqnarray}
M_T(H^\pm)=\sqrt{(E_T(\pi)+\cancel{E}_T)^2-(\vec{p}_T(\pi)+\vec{\cancel{p}}_T)^2}
\end{eqnarray}
and the invariant mass of two $b$-jets $M_{bb}$, respectively, as shown in Figs.~\ref{h3hpm-dis} (c) and (d).
We thus apply the following kinematic cuts
\begin{eqnarray}
&&\cancel{E}_T>M_{H^\pm}/3, \ p_T^{\rm max}(b)>M_{A^0}/2, \nonumber \\
&&p_T(\pi)>M_{H^\pm}/10+40 \ {\rm GeV}, \ |M_{bb}-M_{A^0}|<M_{A^0}/10.
\end{eqnarray}
The cut efficiencies of the signal and backgrounds after imposing the above cuts are summarized in Table~\ref{cuteff-h3hpm}.
One can see that all the SM backgrounds could be suppressed sufficiently and we expect to achieve good signal significance although our signal is induced by a pure electroweak process.

\begin{table}[tb]
\begin{center}
\begin{tabular}{|c|c|c|c|c|c|}
\hline
cut efficiencies  & basic cuts & $p_T^b$ & $\cancel{E}_T$ & $p_T^\pi$ & $M_{bb}$
\\ \hline
$H^\pm A^0 (300)$ & 0.67 & 0.64 & 0.55 & 0.41  & 0.38   \\
$H^\pm A^0 (800)$ & 0.86 & 0.81 & 0.68 & 0.57  & 0.55   \\
\hline \hline
$bbW$ (300) & 0.0064 & 0.00093 & 0.00057 & $0.00017$ & $1.5\times 10^{-5}$   \\
$bbW$ (800) & 0.0064 & $4.0\times 10^{-5}$ & $2.5\times 10^{-5}$ & $5.2\times 10^{-6}$ & negligible   \\
\hline
$bt$ (300) & 0.072 & 0.021 & 0.011  & 0.0017  & $1.8\times 10^{-4}$   \\
$bt$ (800) & 0.072 & 0.0024 & 0.001  & 0.0001  & $2.4\times 10^{-5}$   \\
\hline
$Wg$ (300) & 0.011 & 0.0021 & 0.0012  & 0.00022  & $3.2\times 10^{-5}$ \\
$Wg$ (800) & 0.011 & 0.00012 & $5.6\times 10^{-5}$  & $8.5\times 10^{-6}$  & $7.5\times 10^{-7}$ \\
\hline
$t\bar{t}$ (300) & 0.004 & 0.0006 & 0.00029  & $4.3\times 10^{-5}$  & $9.5\times 10^{-6}$ \\
$t\bar{t}$ (800) & 0.004 & $5.5\times 10^{-6}$ & $1.8\times 10^{-6}$  & $2.5\times 10^{-7}$  & negligible \\
\hline
\end{tabular}
\end{center}
\caption{The cut efficiencies for $pp\to H^{\pm}A^{0}\to \tau^\pm \nu_\tau b\bar{b}$ and the SM backgrounds after consecutive cuts with $\tau^\pm\to \pi^\pm \nu_\tau$ channel at the 27 TeV LHC. We take $M_{H^{\pm}}=M_{A^0}=300$ or 800 GeV.}
\label{cuteff-h3hpm}
\end{table}

\begin{figure}[tb]
\begin{center}
\minigraph{7cm}{-0.05in}{(a)}{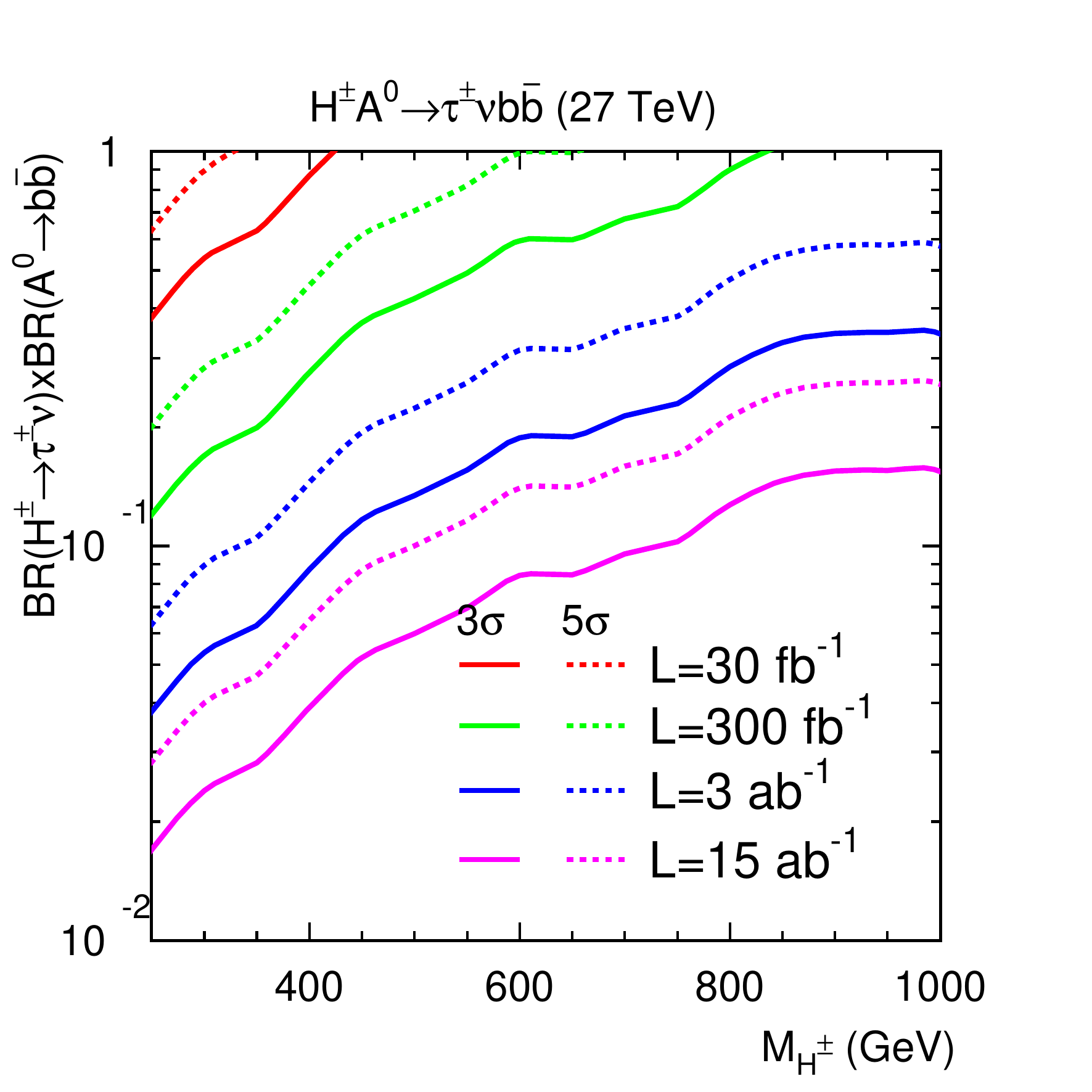}
\minigraph{6.5cm}{-0.05in}{(b)}{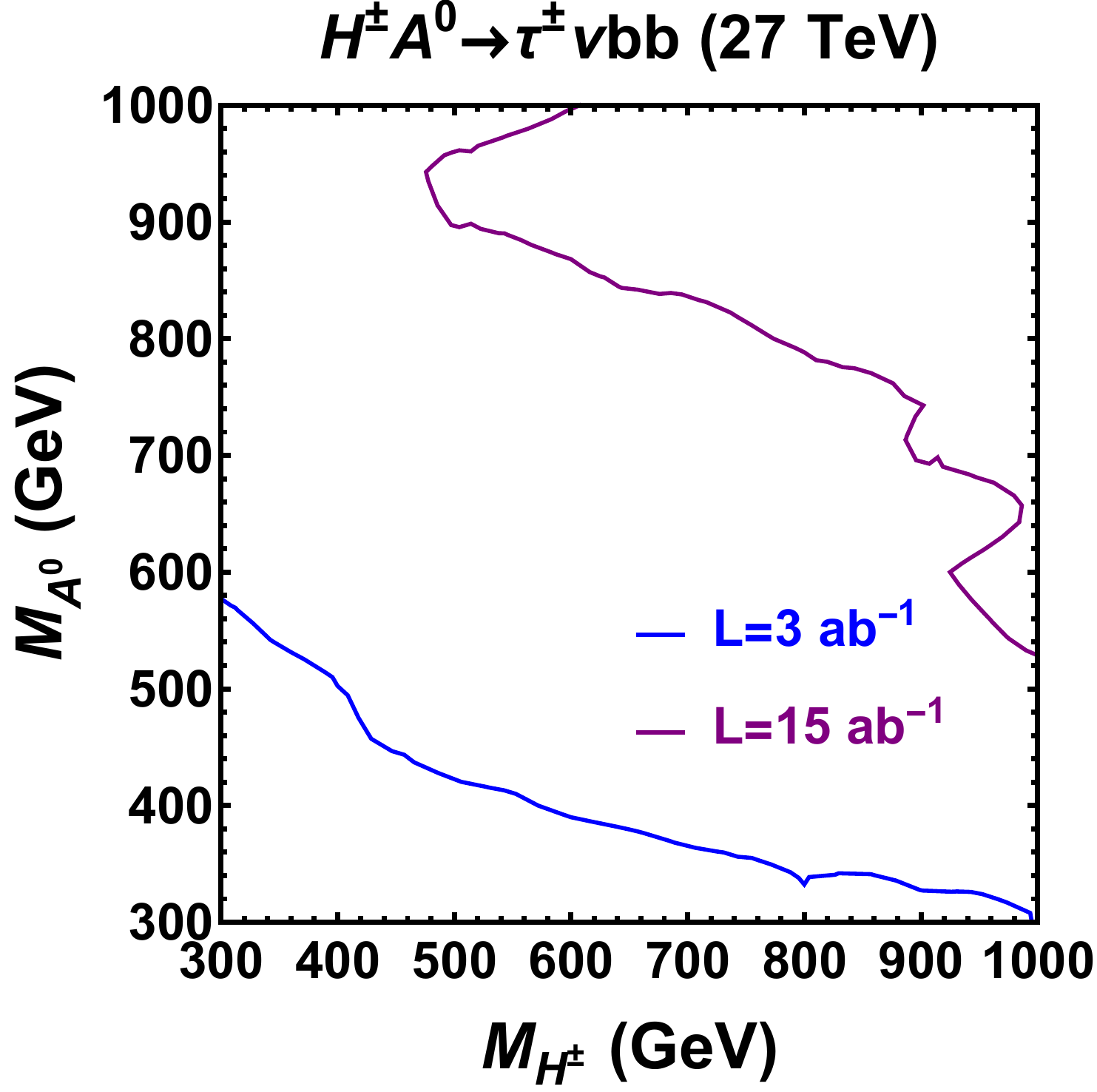}
\end{center}
\caption{
Left: Reach of ${\rm BR}(H^\pm\to \tau^\pm \nu_\tau)\times {\rm BR}(A^0\to b\bar{b})$ versus $M_{H^\pm}$ for $pp\to H^{\pm}A^{0}\to \tau^\pm \nu_\tau b\bar{b}$.
We assume $M_{A^0}=M_{H^\pm}$.
Right: Discovery contour in the plane of $M_{A^0}$ versus $M_{H^\pm}$. We assume ${\rm BR}(H^\pm\to \tau^\pm \nu_\tau)\times {\rm BR}(A^0\to b\bar{b})=20\%$.}
\label{sigma-h3hpm}
\end{figure}

As the $H^\pm A^0$ production is independent of any model parameters, except for the Higgs masses, the only unknown in our signal process can be extracted as the decay branching fractions of $H^\pm$ and $A^0$. In Fig.~\ref{sigma-h3hpm} (a) we show the reach of the product of branching fractions, i.e. ${\rm BR}(H^\pm\to \tau^\pm \nu_\tau)\times {\rm BR}(A^0\to b\bar{b})$, with degenerate spectrum $M_{A^0}=M_{H^\pm}$ and different luminosity assumptions. For $M_{A^0}=M_{H^\pm}\simeq 300$ GeV, with 15 ab$^{-1}$ luminosity, the discovery limit of the branching fraction product can be as small as $3\times 10^{-2}$. With ${\rm BR}(H^\pm\to \tau^\pm \nu_\tau)\times {\rm BR}(A^0\to b\bar{b})=20\%$, the maximal discovery mass of degenerate heavy Higgs bosons are around 450 GeV and 800 GeV with an integrated luminosity of 3 ab$^{-1}$ and 15 ab$^{-1}$, respectively. We also vary the masses of the charged Higgs and the CP-odd Higgs and display the discovery region with respect to the two masses in Fig.~\ref{sigma-h3hpm} (b), by fixing the branching fraction product to be $20\%$. The regions to the left of the curves can be covered by 5$\sigma$ discovery.

\subsection{$H^\pm A^0\to t\bar{b}(\bar{t}b) b\bar{b}$}
Next we study the signal induced by $H^\pm \to tb$ with the top quark's leptonic decay, i.e. $H^\pm A^0\to t\bar{b}(\bar{t}b) b\bar{b}\to bbbb\ell^\pm \nu$, and the leading SM backgrounds including
\begin{itemize}
\item the virtual $W$ process: $q\bar{q}'\to gW^{\pm\ast}\to t\bar{b}(\bar{t}b) b\bar{b}$ ,
\item $tb$ production: $q\bar{q}'\to W^{\pm\ast}\to gt\bar{b}(\bar{t}b)\to t\bar{b}(\bar{t}b) b\bar{b}$ .
\end{itemize}
As we require the CP-odd Higgs to decay into $b\bar{b}$, this case still has the Jacobian peak around $p_{Tb}\sim M_{A^0}/2$.
The missing transverse energy here is softer than that in $H^\pm A^0\to \tau^\pm\nu b\bar{b}$ mode as the neutrino is from the subsequent decay of top quark.
Thus, we apply the following kinematic cuts in addition to the basic acceptance cuts described in Sec.~\ref{sec:singleHA} and \ref{sec:singleHpm}.
\begin{eqnarray}
\cancel{E}_T>40 \ {\rm GeV}, \ p_T^{\rm max}(b)>M_{A^0}/2.
\end{eqnarray}

\begin{figure}[h!]
\begin{center}
\minigraph{7cm}{-0.05in}{(a)}{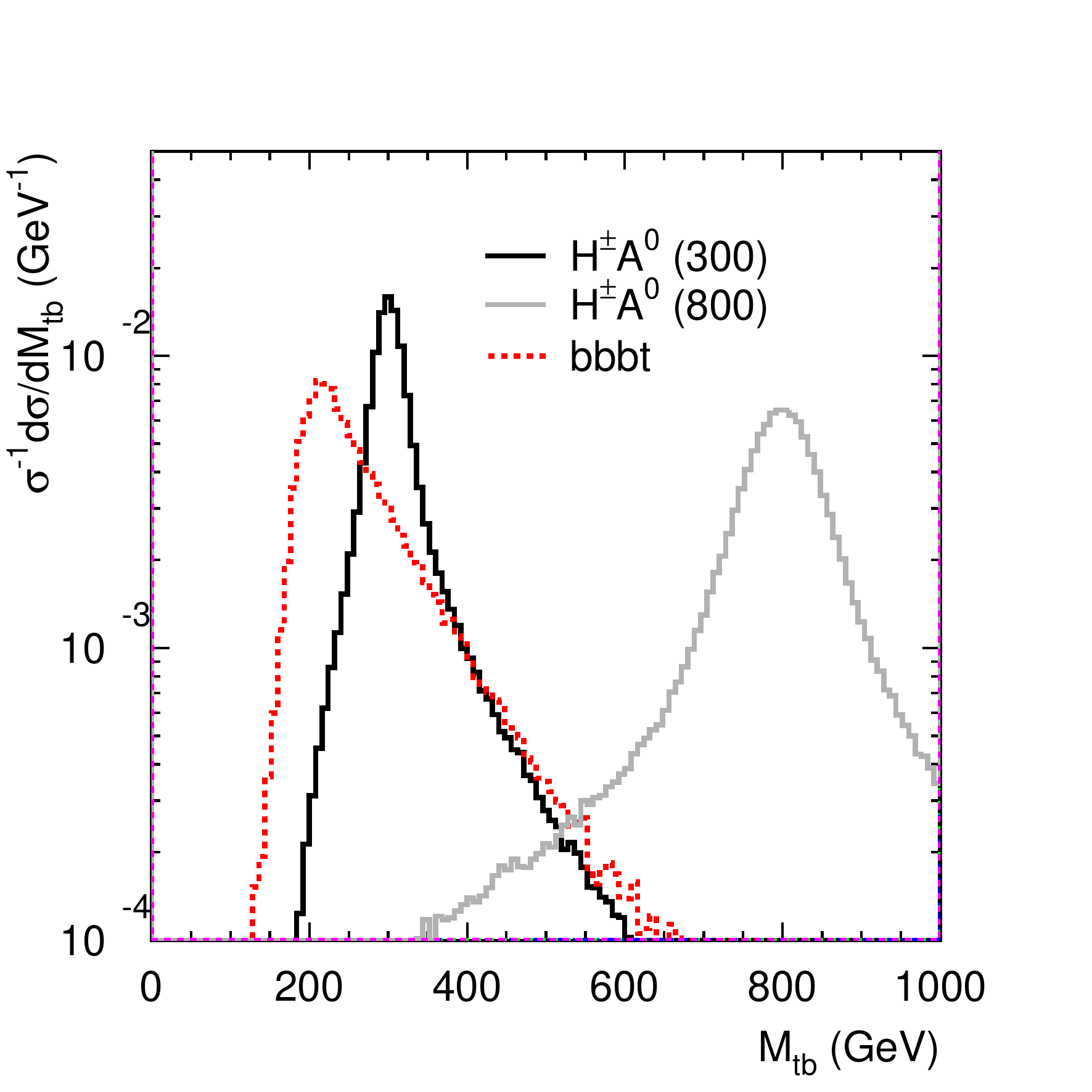}
\minigraph{7cm}{-0.05in}{(b)}{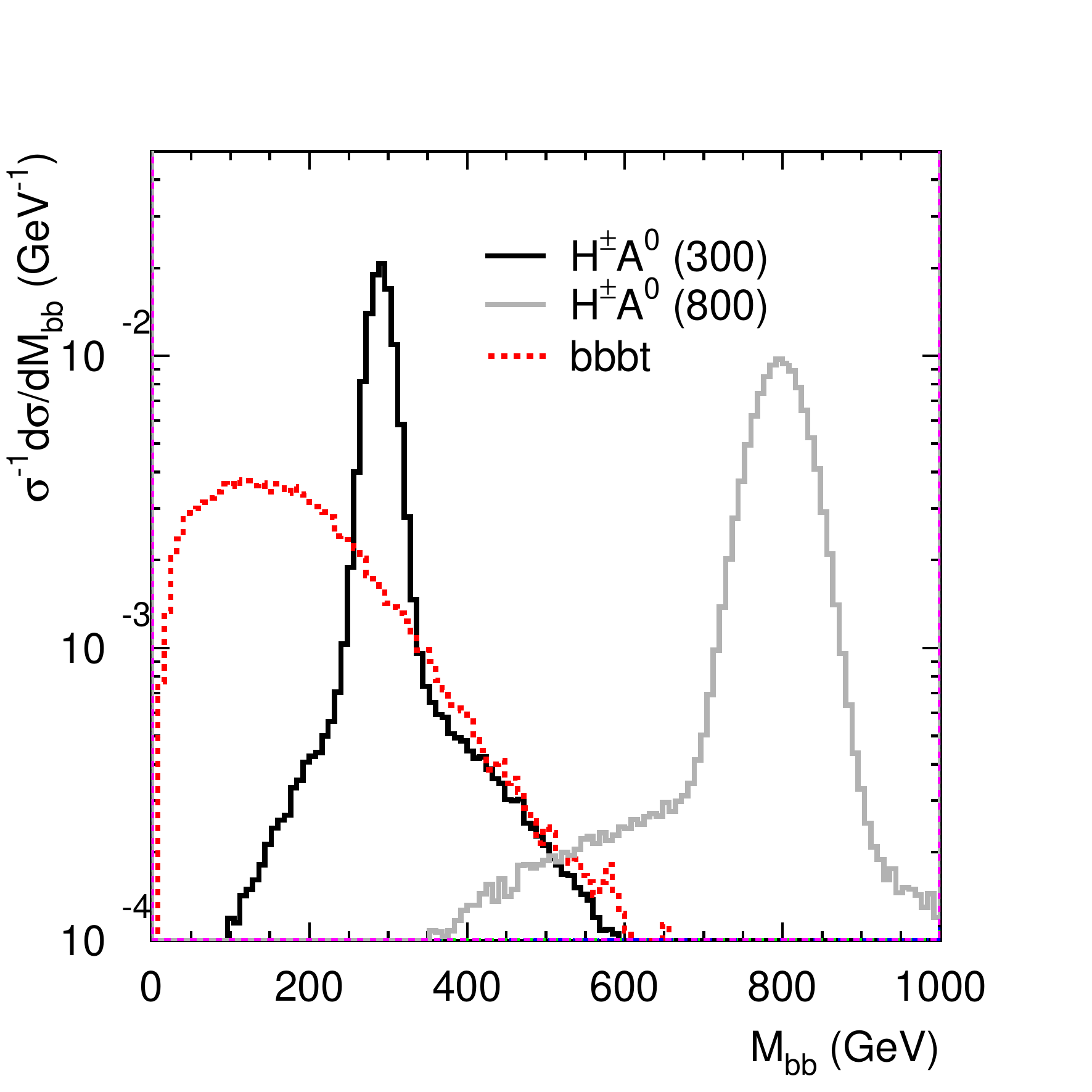}\\
\minigraph{7cm}{-0.05in}{(c)}{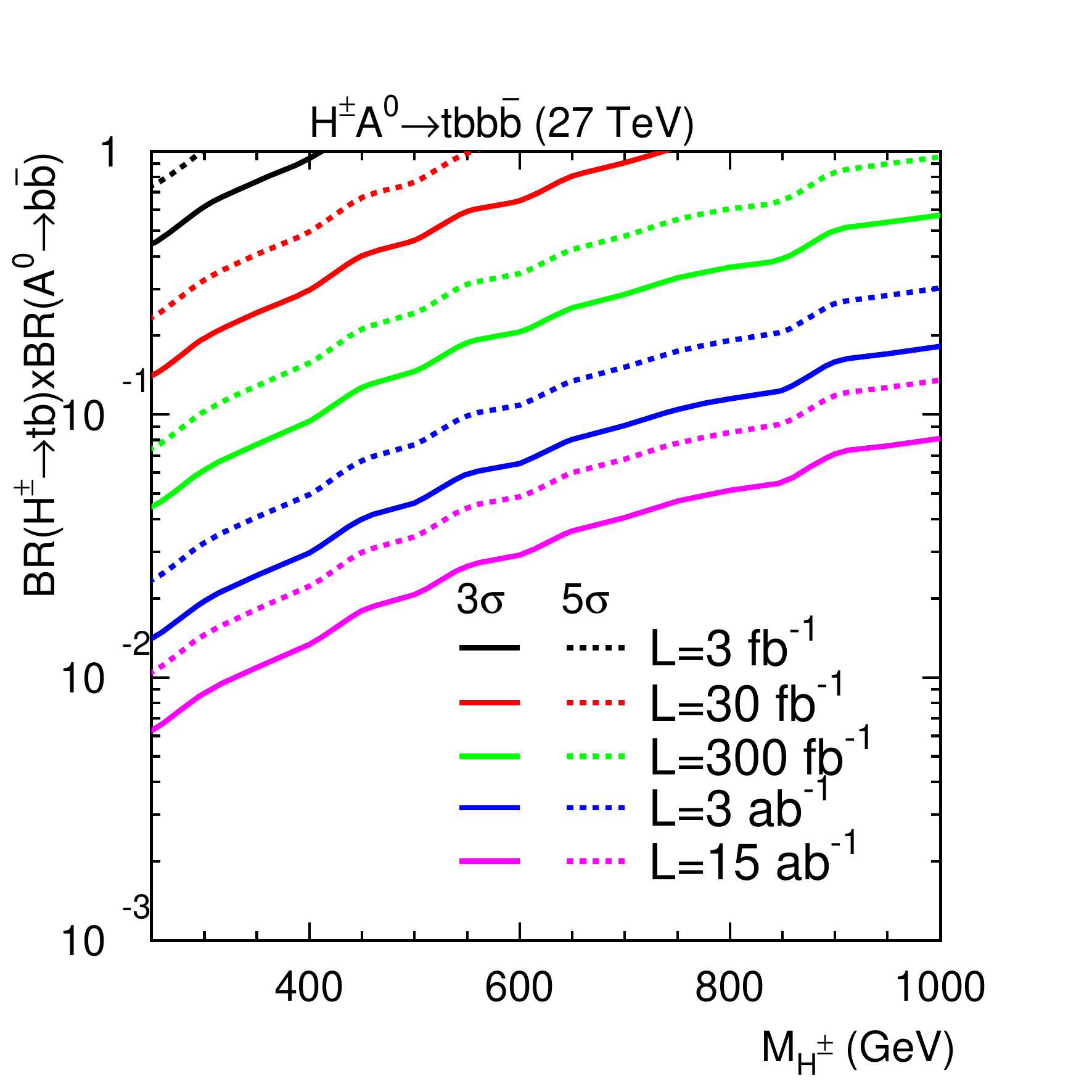}
\end{center}
\caption{Top: The differential cross section distributions of $M_{tb}$ (a) and $M_{bb}$ (b) for the signal $pp\to H^{\pm}A^{0}\to t\bar{b}(\bar{t}b) b\bar{b}\to bbbb\ell^\pm \nu$ and backgrounds at the 27 TeV LHC. Bottom: Reach of ${\rm BR}(H^\pm\to tb)\times {\rm BR}(A^0\to b\bar{b})$ versus $M_{H^\pm}$ for $pp\to H^{\pm}A^{0}\to t\bar{b}(\bar{t}b) b\bar{b}\to bbbb\ell^\pm \nu$, assuming $M_{A^0}=M_{H^\pm}$.}
\label{h3hpmbbbt-dis}
\end{figure}

The leptonic $W$ boson from the top quark can be reconstructed using the method described in Sec.~\ref{sec:Htb}.
Because of the complexity from the four $b$-jets in our signal, when requiring the correct combination to reconstruct $M_{H^\pm}$ and $M_{A^0}$,
we assume and make use of the nearly-equal mass spectrum of $H^\pm$ and $A^0$. The obtained invariant masses of $tb$ and $b\bar{b}$ are shown in Figs.~\ref{h3hpmbbbt-dis} (a) and (b), respectively. Then, we can take two mass windows near the resonances
\begin{eqnarray}
|M_{tb}-M_{H^\pm}|<M_{H^\pm}/10, \ \ |M_{bb}-M_{A^0}|<M_{A^0}/10.
\end{eqnarray}
The cut efficiencies are illustrated in Table~\ref{cuteff-h3hpmbbbt}.

\begin{table}[tb]
\begin{center}
\begin{tabular}{|c|c|c|c|c|c|}
\hline
cut efficiencies  & basic cuts & $p_T^b$ & $\cancel{E}_T$ & $M_{tb}$ & $M_{bb}$
\\ \hline
$H^\pm A^0 (300)$ & 0.34 & 0.33 & 0.25 & 0.16  & 0.14   \\
$H^\pm A^0 (800)$ & 0.45 & 0.43 & 0.39 & 0.27  & 0.26   \\
\hline \hline
$bbbt$ (300) & 0.032 & 0.016 & 0.012 & 0.0025 & 0.00048   \\
$bbbt$ (800) & 0.032 & 0.0024 & 0.0021 & 0.00011 & $1.9\times 10^{-5}$   \\
\hline
\end{tabular}
\end{center}
\caption{The cut efficiencies for $pp\to H^{\pm}A^{0}\to t\bar{b}(\bar{t}b) b\bar{b}\to bbbb\ell^\pm \nu$ and the SM backgrounds after consecutive cuts at the 27 TeV LHC. We take $M_{H^{\pm}}=M_{A^0}=300$ or 800 GeV.}
\label{cuteff-h3hpmbbbt}
\end{table}

In our signal process, the only dependence is again the product of decay branching fractions which is ${\rm BR}(H^\pm\to tb)\times {\rm BR}(A^0\to b\bar{b})$ here. As shown in Fig.~\ref{h3hpmbbbt-dis} (c), with degenerate spectrum $M_{A^0}=M_{H^\pm}\simeq 300$ GeV and 15 ab$^{-1}$ luminosity, the reach of the branching fraction product extends low to the level of $10^{-2}$. With ${\rm BR}(H^\pm\to tb)\times {\rm BR}(A^0\to b\bar{b})=10\%$, the heavy Higgs bosons with 600 GeV and 900 GeV of mass can be discovered with an integrated luminosity of 3 ab$^{-1}$ and 15 ab$^{-1}$, respectively.

\subsection{$H^+H^-\to \tau^+\tau^-\nu\bar{\nu}, t\bar{b}\bar{t}b$}
The first signal of $H^+H^-$ pair production consists of two tau leptons plus missing energy $H^+H^-\to \tau^+\tau^- \nu_\tau \bar{\nu}_\tau$, followed by $\tau^\pm\to \pi^\pm \nu$. The irreducible SM backgrounds are from diboson productions
\begin{eqnarray}
W^+W^-\to \tau^+\nu_\tau\tau^-\bar{\nu}_\tau, \ ZZ\to \tau^+\tau^-\nu\bar{\nu},
\end{eqnarray}
and the reducible contribution is
\begin{eqnarray}
W^\pm Z\to \tau^+\tau^-\ell^\pm \nu_\ell
\end{eqnarray}
which can also be vetoed by the requirement in Eq.~(\ref{veto}).

\begin{figure}[h!]
\begin{center}
\minigraph{7cm}{-0.05in}{(a)}{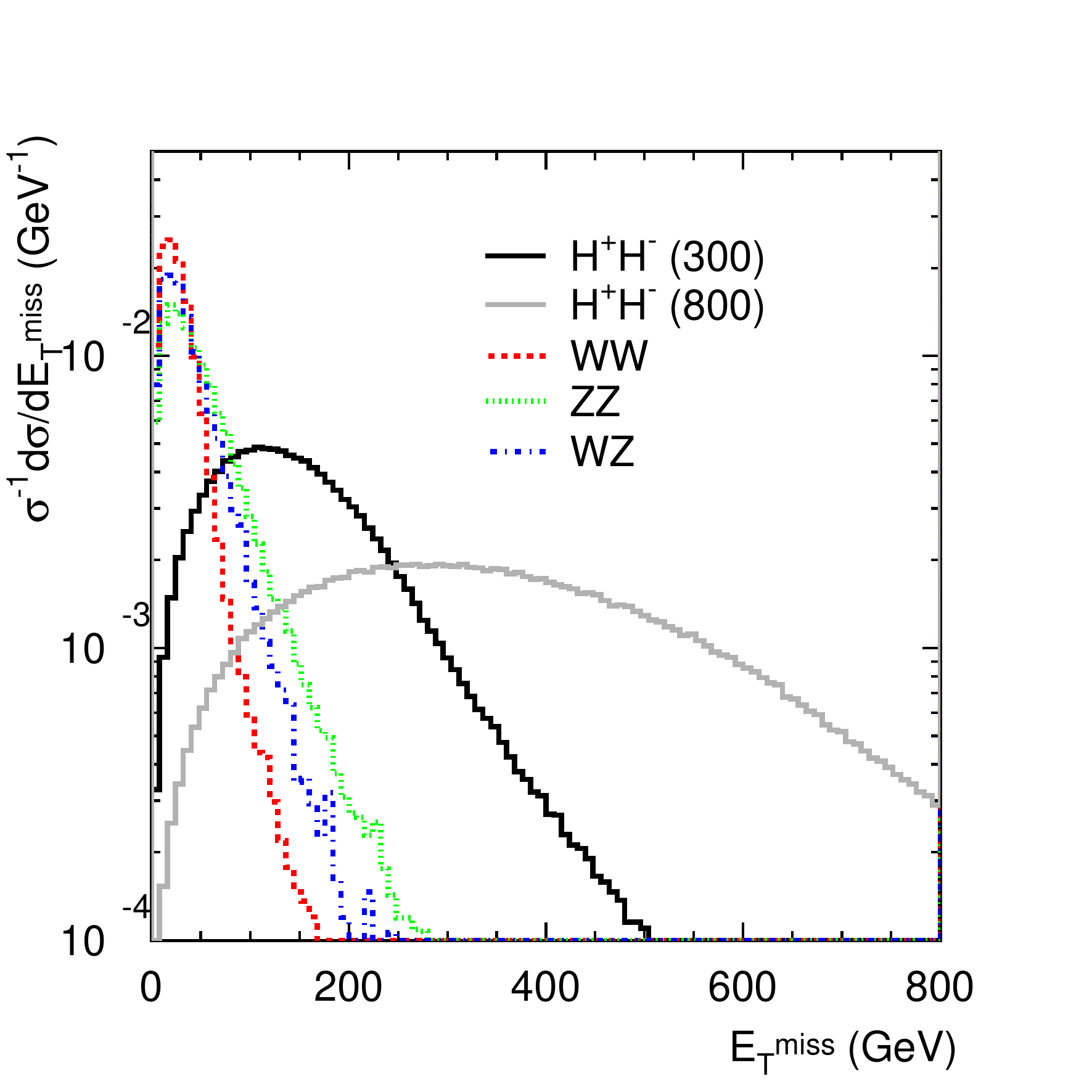}
\minigraph{7cm}{-0.05in}{(b)}{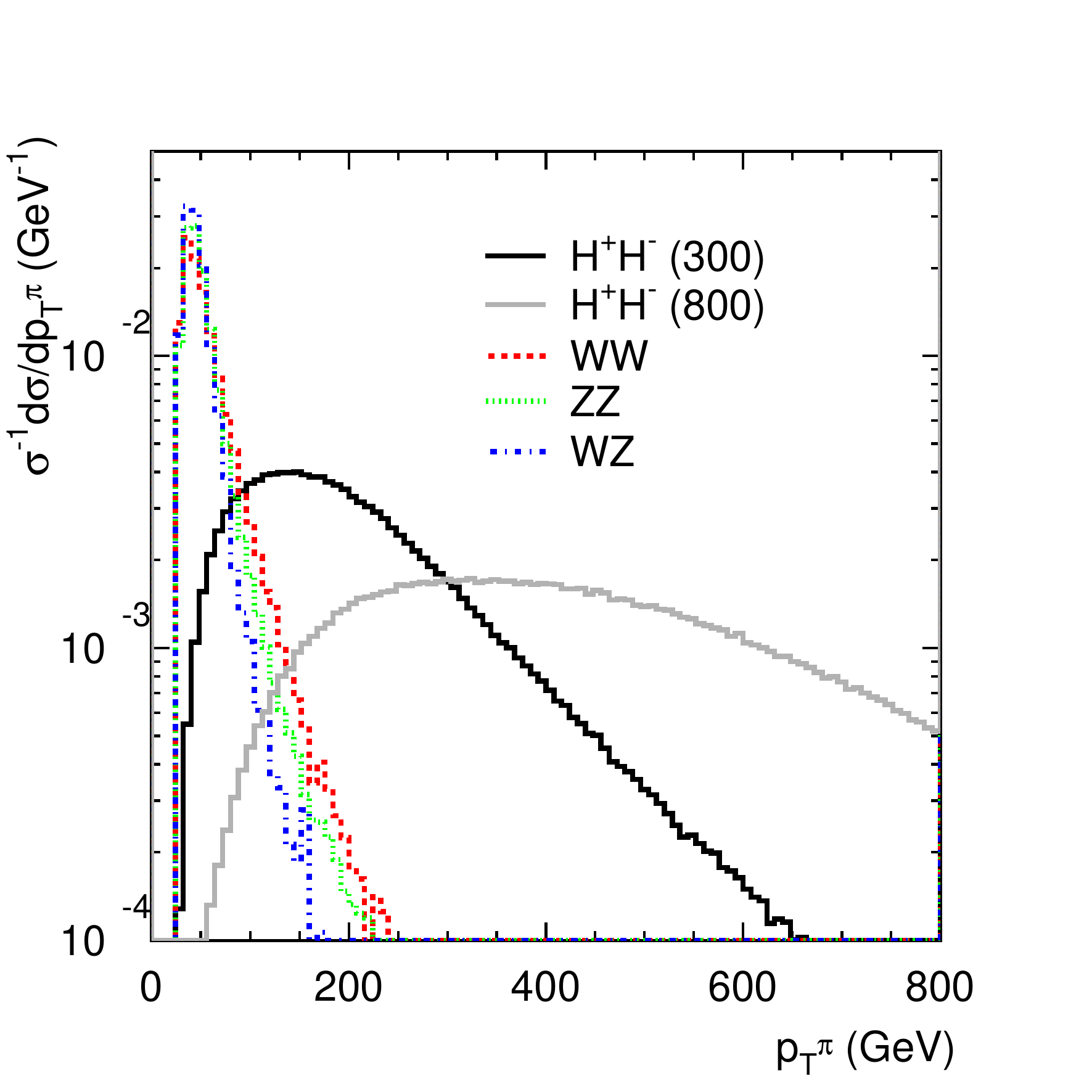}\\
\minigraph{7cm}{-0.05in}{(c)}{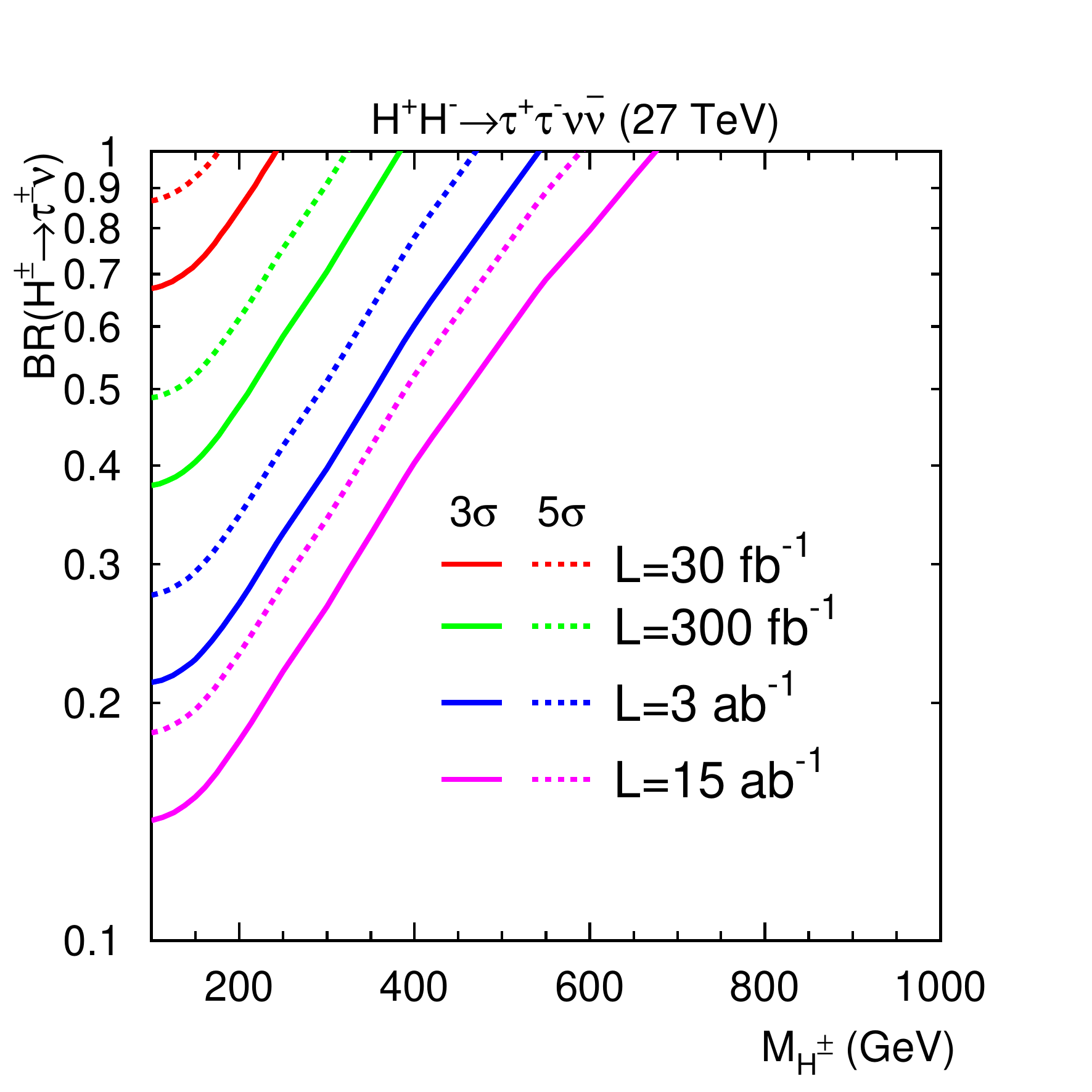}
\end{center}
\caption{Top: The differential cross section distributions of $\cancel{E}_T$ (a) and $p_T(\pi)$ (b) for the signal $pp\to H^+H^-\to \tau^+\tau^- \nu_\tau \bar{\nu}_\tau$ and backgrounds at the 27 TeV LHC. Bottom: Reach of ${\rm BR}(H^\pm\to \tau^\pm \nu)$ versus $M_{H^\pm}$ for $pp\to H^+H^-\to \tau^+\tau^- \nu_\tau \bar{\nu}_\tau$.}
\label{h+h--dis}
\end{figure}

The distributions of signal and backgrounds at the
27 TeV LHC after the basic cuts are shown in Fig.~\ref{h+h--dis}, for (a) missing transverse energy $\cancel{E}_T$ and (b) transverse pion momentum $p_T(\pi)$.
One can see that the tau polarization effect mentioned above tends to be more dramatic in this channel (in comparison
with the $WW$ background). We thus strengthen the missing energy and $p_T(\pi)$ as follows
\begin{eqnarray}
\cancel{E}_T> 100 \ {\rm GeV}, \ \ \ p_T^{\rm max}(\pi)> 100 \ {\rm GeV} .
\end{eqnarray}
Cut efficiencies are collected in Table~\ref{cuteff-hphm}. Due to the missing neutrinos from both the charged Higgs and the tau lepton in this channel, one is unable to reconstruct the charged Higgs boson or build a transverse mass to estimate the signal observability. The signal-to-background ratio is not expected to be improved as much as the associated production analyzed in Sec.~\ref{HpmAtau}. Figure~\ref{h+h--dis} (c) shows the reach of ${\rm BR}(H^\pm\to \tau^\pm \nu)$ versus $M_{H^\pm}$ for $pp\to H^+H^-\to \tau^+\tau^- \nu_\tau \bar{\nu}_\tau$. One can see that this channel can access the decay branching fraction to be $20\%$ for the charged Higgs just above the top quark threshold with 15 ab$^{-1}$ luminosity.

\begin{table}[tb]
\begin{center}
\begin{tabular}{|c|c|c|c|c|}
\hline
cut efficiencies  & basic cuts & $\cancel{E}_T$ & $p_T^\pi$ 
\\ \hline
$H^+ H^- (300)$ & 0.7 & 0.49 & 0.46 \\ 
$H^+ H^- (800)$ & 0.89 & 0.84 & 0.84 \\ 
\hline \hline
$WW$ & 0.024 & 0.00056 & 0.00056 \\ 
\hline
$ZZ$ & 0.084 & 0.011 & 0.0052 \\ 
\hline
$WZ$ & 0.0094 & 0.00062 & 0.00026 \\ 
\hline
\end{tabular}
\end{center}
\caption{The cut efficiencies for $pp\to H^+H^-\to \tau^+\tau^- \nu_\tau \bar{\nu}_\tau$ and the SM backgrounds after consecutive cuts with $\tau^\pm\to \pi^\pm \overset{(-)}{\nu_\tau}$ channel at the 27 TeV LHC. We take $M_{H^{\pm}}=300$ or 800 GeV.}
\label{cuteff-hphm}
\end{table}

Finally, we consider semi-leptonic channel $H^+H^-\to t\bar{b}\bar{t}b\to bbbbjj\ell^\pm\nu$ induced by $H^\pm\to tb$ and the leading SM background $b\bar{b}t\bar{t}$. Using the methods mentioned in Sec.~\ref{sec:Htb}, the two charged Higgses can be fully reconstructed. The sensitivity of this search is limited by the efficiency of the top quark tagging due to smaller typical transverse momenta.
Assuming BR$(H^\pm \to tb)=1$, we can accumulate 250 (9) signal events for $M_{H^\pm}=300 \ (800)$ GeV with 15 ab$^{-1}$ luminosity. To discover the charged Higgs with the mass of 300 GeV, one needs 50 ab$^{-1}$ luminosity. This mode is thus not optimistic for probing the charged Higgs.

\section{Conclusions}
\label{sec:Concl}

New Higgs bosons are present in many of new physics models and their direct searches yield no signal observation in the LHC experiments so far. LHC upgrades with higher energy, such as the HE-LHC and FCC-hh, are thus motivated to carry out the search for heavy non-SM Higgs bosons.

In this paper, we investigate the discovery potential of the HE-LHC with 27 TeV C.M.~energy for the heavy Higgses in Type-II 2HDM.
To accommodate the theoretical bounds and experimental limits, we assume degenerate Higgs spectrum $M_{H^0}\approx M_{A^0}\approx M_{H^\pm}$ and the parameter $\cos(\beta-\alpha)$ near the alignment limit. We analyze the typical production and decay modes of non-SM Higgses and present the implication on the parameter space of Type-II 2HDM.

We explore the observability of the heavy neutral Higgs bosons by examining the leading decay channel $H^0/A^0\to \tau^+\tau^-$ and the clean signals from $H^0\to W^+W^-, ZZ$ via gluon-gluon fusion production. With realistic decay branching fractions, for $\tan\beta\sim 1$, the 27 TeV LHC can probe the neutral Higgs as heavy as 2 TeV with the luminosity of 15 ab$^{-1}$. For large values of $\tan\beta (\sim 50)$, the $\tau\tau$ channel gives the better sensitivity and can reach heavy Higgs mass up to 1.1 TeV. For the charged Higgs bosons, we consider the inclusive process with the charged Higgs produced in association with a top quark that is $gb\to t H^\pm$. The region below $\tan\beta\sim 1$ can not be covered by $5\sigma$ discovery of $H^\pm\to \tau^\pm \nu$ decay mode due to the suppression of the decay branching fraction. The final states with $t H^\pm\to bt\bar{t}$ prove to be a very sensitive channel for regions with both small and large $\tan\beta$.
For $\tan\beta\sim 1 \ (50)$, the $bt\bar{t}$ channel can extend the reach to about $M_{H^\pm}\approx 2 \ (1.4)$ TeV with 300 fb$^{-1}$ luminosity.

The electroweak productions of non-SM Higgs boson pairs provide complementary signals in the determination of the nature of the Higgs sector.
They are benefitted from pure electroweak gauge interactions and independent of additional model parameters except for Higgs masses.
We explore the pair productions $H^\pm A^0$ and $H^+ H^-$, followed by $H^\pm \to \tau^\pm \nu, tb$ and $A^0\to b\bar{b}$ decays.
With ${\rm BR}(H^\pm\to \tau^\pm \nu_\tau, tb)\times {\rm BR}(A^0\to b\bar{b})=(10-20)\%$, the maximal discovery mass of degenerate heavy Higgs bosons is around $800-900$ GeV with an integrated luminosity of 15 ab$^{-1}$. The $pp\to H^+ H^-$ production is not optimistic to probe the charged Higgs.
The $pp\to H^+H^-\to \tau^+\tau^-\nu\bar{\nu}$ channel can access the decay branching fraction ${\rm BR}(H^\pm\to \tau^\pm \nu_\tau)$ to be $20\%$ for light charged Higgs with 15 ab$^{-1}$ luminosity.

\acknowledgments
We would like to thank Tao Han for collaboration at the early stage of this project and valuable discussions.
This work is supported by ``the Fundamental Research Funds for the Central Universities'', Nankai University (Grant Number 63191522, 63196013).


\bibliography{refs}

\end{document}